\documentclass[%
 reprint,showpacs,
 amsmath,amssymb, 
 aps,nofootinbib,superscriptaddress
]{revtex4-1}
    \usepackage{float}
\usepackage{graphicx, xcolor}% Include figure files
\usepackage{dcolumn}% Align table columns on decimal point
\usepackage[colorlinks=true, linkcolor = red, citecolor = blue]{hyperref}
\usepackage{subcaption}
\usepackage{amsmath}
\usepackage[normalem]{ulem}
\usepackage{soul}

\usepackage{float}
\usepackage{xcolor}

\newcommand{\be}{\begin{equation}}  \newcommand{\ee}{\end{equation}}
\newcommand{\bea}{\begin{eqnarray}} \newcommand{\eea}{\end{eqnarray}}

\DeclareMathOperator{\erf}{erf}
\DeclareMathOperator{\erfc}{erfc}

\begin{document}
\newcommand{\lu}[1]{\textcolor{red}{#1}}
\newcommand{\quita}[1]{\textcolor{orange}{#1}}
\newcommand{\JCH}[1]{\textcolor{blue}{#1}}
\newcommand{\TDG}[1]{\textcolor{green}{#1}}
\title{Primordial black hole formation during slow-reheating: A review}% Force line breaks with \\

\author{Luis E.~Padilla}\thanks{On leave at Astronomy Unit, Queen Mary University of London, United Kingdom.}
\email{l.padilla@qmul.ac.uk}
\affiliation{Instituto de Ciencias Físicas, Universidad Nacional Autónoma de México, 62210, Cuernavaca, Morelos, México.}

\author{Juan Carlos Hidalgo}
  \email{hidalgo@icf.unam.mx}
\affiliation{Instituto de Ciencias Físicas, Universidad Nacional Autónoma de México, 62210, Cuernavaca, Morelos, México.}
  \author{Tadeo D.~Gomez-Aguilar}
  \email{tadeo.dga@icf.unam.mx}
  \affiliation{Instituto de Ciencias Físicas, Universidad Nacional Autónoma de México, 62210, Cuernavaca, Morelos, México.}
  \author{Karim A.~Malik}
  \email{k.malik@qmul.ac.uk}
  \affiliation{Astronomy Unit, Queen Mary University of London, London, E1 4NS, United Kingdom.}
\author{Gabriel German}
  \email{gabriel@icf.unam.mx}
  \affiliation{Instituto de Ciencias Físicas, Universidad Nacional Autónoma de México, 62210, Cuernavaca, Morelos, México.}

\date{\today}% 

\begin{abstract}
In this paper we review the possible mechanisms for the production of primordial black holes (PBHs) during a slow-reheating period {in which the energy transfer of the inflaton field to standard model particles becomes effective at slow temperatures}, offering a comprehensive examination of the theoretical foundations and conditions required for each of formation channel. In particular, we focus on post-inflationary scenarios where there are no self-resonances and the reheating epoch can be described {by the inflaton evolving in} a quadratic-like potential. In the hydrodynamical interpretation of this field during the slow-reheating epoch, the gravitational collapse of primordial fluctuations is subject to conditions on their sphericity, limits on their spin, as well as a maximum velocity dispersion. We show how to account for all conditions and show that PBHs form with different masses depending on the collapse mechanism. Finally we show, through an example, how PBH production serves to probe both the physics after primordial inflation, as well as the primordial powerspectrum at the smallest scales.
\end{abstract}

\maketitle
\section{Introduction}

While the existence of supermassive and stellar-mass black holes today
is thoroughly demonstrated, the  possibility
of a third species of black holes has been hypothesized in recent decades: the formation of primordial black holes
(PBHs) during the early stages of evolution of the universe (see
e.g.~Refs.~\citep{Green:2020jor,Carr:2020gox, Ozsoy:2023ryl, Khlopov:2008qy, Garcia-Bellido:2017fdg, Sasaki:2018dmp, Carr:2020xqk, Escriva:2022duf} for recent reviews). These objects may be key to understand fundamental aspects of
cosmology and particle physics.

One epoch in which primordial black holes might have emerged is the
reheating phase that immediately followed the period of cosmic
inflation (see
e.g.~Refs.~\citep{doi:10.1146/annurev.nucl.012809.104511,Amin:2014eta}). The
inflationary paradigm postulates that the early universe experienced a period of rapid and exponential expansion in its earliest moments (see
e.g.~Refs.~\citep{liddle2000cosmological,LYTH19991,OLIVE1990307, Linde:1984ir, Martin:2018ycu, Baumann:2009ds,Odintsov:2023weg}). This
inflationary period played a crucial role in explaining the observed
flatness of the universe and the uniformity of the cosmic microwave
background radiation, also offering an elegant explanation for the
large-scale structure of the universe (see e.g.~Ref.~
\citep{Vazquez:2018qdg,Langlois:2010xc}). After inflation, the universe
entered a reheating phase characterized by the decay of the inflaton
field, resulting in the transfer of its energy to matter
and radiation. This process eventually led to the emergence of a hot
and dense environment, providing the necessary conditions for the
subsequent stages of cosmic evolution.

The process of reheating represents a crucial yet relatively understudied chapter in cosmology. It not only determines the thermal properties of the early universe but also plays a fundamental role in particle production and the subsequent formation of cosmic structures (see e.g.~Refs.~\citep{doi:10.1146/annurev.nucl.012809.104511,ElBourakadi:2021blc,Amin:2014eta,Allahverdi:2010xz}). Among the various possibilities that arise during reheating, the formation of PBHs has garnered significant attention due to their particular characteristics and potential cosmological implications (see e.g.~\citep{Padilla:2021zgm,Hidalgo:2022yed,Padilla:2023lbv,DeLuca:2021pls,Hidalgo:2017dfp,Harada:2022xjp,Bhattacharya:2023ztw,Martin_2020,Carr:2018nkm,Harada:2016mhb,Harada:2017fjm,Carr:2017edp,Carrion:2021yeh}). 

%For example, in \citep{Gomez-Aguilar:2023bej} it has been shown that different constraints from cosmological and astrophysical observations could constrain the primordial power spectrum more strongly than in the standard scenario of PBH formation during radiation.

A key aspect of PBH formation during reheating lies in the collapse threshold for the density contrast compared to the formation process in a radiation-dominated
background. During the inflationary phase the rapid expansion of space
stretches small-scale quantum fluctuations to macroscopic and
cosmological scales. Once inflation ceases and scales slowly re-enter the horizon, these fluctuations undergo substantial 
growth due to the dynamics of the reheating process. Consequently,
localized regions with remarkably high density emerge. If the density
within these regions surpasses a critical threshold, they may collapse and
form PBHs.

Understanding the formation and characteristics of PBHs during the
reheating phase poses a challenge spanning the
fields of cosmology, particle physics, and astrophysics. By
investigating the mechanisms of PBH formation, their mass spectrum,
and cosmological abundance, we may gain insights into the
fundamental physics that governed the early universe. Moreover, such 
PBHs hold potential in elucidating intriguing astrophysical phenomena,
including dark matter (see e.g.~\citep{Frampton:2010sw,Ivanov:1994pa,Carr:2020xqk,Carr:2021bzv,10.3389/fspas.2021.681084, Green:2020jor}),
gravitational waves (see e.g.~\citep{Franciolini:2021nvv,Ballesteros:2022hjk,papanikolaou:hal-02999527,Eroshenko_2018,refId0,Papanikolaou_2023}),
and the origins of supermassive black holes (see
e.g.~\citep{Dolgov:2019vlq,Kawasaki:2012kn,Duechting:2004dk,Bernal_2018}). All these possibilities further emphasize  the significance of PBHs in our quest
for a comprehensive understanding of the universe.

In this paper we intend to present a comprehensive review of the formation of
PBHs during the reheating epoch. Starting with the theoretical foundations,
we discuss the various formation mechanisms. By studying the details of PBH formation during reheating, we aim to
contribute to the understanding of the early universe and shed light
on the nature of black holes, providing potential avenues
for future research and novel insights into the cosmic dark sector.

\section{Inflation, preheating and the reheating epochs}

\subsection{Inflation setting the initial conditions for reheating}

\begin{figure}[H]
    \centering
    \includegraphics[width=3.3in]{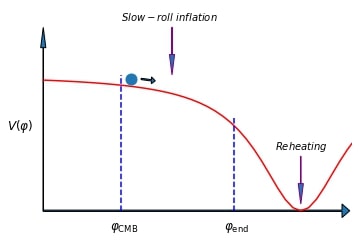}
    \caption{\footnotesize{Schematic inflationary and reheating phases. Initially, the inflaton slowly rolls along its potential until it reaches a critical point where $\epsilon\simeq 1$ at $\varphi \simeq \varphi_{\rm end}$. Subsequently, the inflaton transits rapidly towards the bottom of the potential, where it oscillates rapidly at around the minimum. At such stage the process of reheating takes place.}}
    \label{fig:inflation}
\end{figure}

As mentioned above, cosmological inflation
(see e.g.~ \citep{Vazquez:2018qdg,Langlois:2010xc,PhysRevD.23.347}) refers to a period of accelerated expansion of space. In the framework of general relativity, inflation usually stipulates the existence of a scalar field as the dominant energy content of the universe during this period. In its
simplest form, the inflationary scenario is described by the
action %{\citep{Senatore:2016aui,Vazquez:2018qdg}} 
\begin{equation}
    S = \int d^{4}x\sqrt{-g}\mathcal{L} = \int d^{4}x\sqrt{-g}\left[\frac{1}{2}\partial_{\mu}\varphi\partial^{\mu}\varphi - V(\varphi)\right]\,.
    \label{eq:action}
\end{equation}
For the inflaton $\varphi$ to drive the inflationary epoch, its
energy must be dominated by a nearly
constant potential energy $V(\varphi)$. In this case,
the inflaton field behaves effectively like a cosmological constant,
causing the universe to expand exponentially. When
the kinetic part of the inflaton field is subdominant compared to the
potential part $V(\varphi)$, inflation kicks in, whereas when both
quantities become comparable, the inflationary period ends. This
requirement is typically expressed in the slow-roll conditions
$\epsilon\equiv (1/2)[V'(\varphi)/V(\varphi)]^2\ll 1$ and
$\eta\equiv |V^{''}(\varphi)/V(\varphi)|\ll 1$ for the inflaton to
produce inflation and $\epsilon \simeq 1$ to end the inflationary
epoch (see also Fig.~\ref{fig:inflation} for a sketch of
the inflationary potential).

%We defined the \textit{density contrast} by $\delta\equiv\delta\rho/\rho$. To analyze this, it is convenient to work in Fourier description, this means that quantities are replaced $\delta(x,t)$ becomes $\delta k(k,t)$. Here assume adiabatic initial conditions, which require that matter and radiation perturbations are initially in perfect thermal equilibrium\footnote{In inflationary perturbation calculations, the size of perturbations compared to the comoving Hubble scale is crucial.}. Consequently, the density contrast for different species in the Universe satisfies this condition \citep{Byrnes:2021jka}. 
The scalar perturbations during the inflationary epoch are of special importance since they are attributed the generation of the initial inhomogeneities that gave rise to the large scale structure in the universe. Inflaton fluctuations may also get to form PBHs in the early universe. That is, if an initial perturbation is dense enough when it reenters the cosmological horizon, it can collapse under its own gravity to form a black hole.\footnote{In this review we focus on PBHs formed at horizon entry. For PBHs formed inside the horizon see however Refs.~\citep{Zaballa:2006kh,Lyth:2005ze,Torres-Lomas:2014bua}.} Hence, in order to assess the probability of  formation of PBHs in the early universe, it is necessary to determine precisely the average amplitude of scalar perturbations generated during inflation, that is, the primordial power spectrum. 

The evolution of the scalar field is governed by the Klein-Gordon equation. In order to calculate the amplitude distribution or power spectrum of the field fluctuations $\delta\varphi$, the perturbed Klein-Gordon equation must be solved. Adopting the flat gauge, $\delta\varphi$ is turned into the Sasaki-Mukhanov variable 
(see e.g.~Ref.~\citep{Malik:2008im}) and it proves convenient to define a new variable (in Fourier space)
\begin{equation}
    {u}_{k}\equiv a\delta\varphi_{k}\,,
\end{equation}%
where $k$ is a comoving wavenumber scale. This allows us to rewrite the perturbed Klein-Gordon equation as
the so-called Sasaki-Mukhanov equation \citep{Sasaki:1986hm,Mukhanov:1988jd}:
\begin{equation}
    u_{k}^{''}+\left(k^2-\frac{z^{''}}{z}\right)u_{k} = 0\,,
\end{equation}

\noindent where a prime ( $'$ ) denotes a derivative with respect to the conformal time $d\tau = dt/a$, $z\equiv a\,\dot\varphi_b/H$, and the subscript $b$ is used to refer to background quantities.  
Note that the comoving curvature perturbation $\mathcal{R}_{k}$ is given in terms of the perturbed scalar field (in flat gauge) $\delta\varphi_{k}$ as
\begin{equation}
    \mathcal{R}_{k} = \frac{a'}{a\varphi_b'}\delta\varphi_{k}\,.
\end{equation}
It is in terms of this quantity that  the power spectrum of scalar perturbations is defined. Explicitly, the dimensionless primordial power spectrum of curvature perturbations is 
\begin{equation}
    \mathcal{P}_{\mathcal{R}}(k)\equiv \left.\frac{k^3}{2\pi^2}|\mathcal{R}_{k}|^2\right|_{k\ll aH}.
\end{equation}

The power spectrum is normalized to the amplitudes derived from the CMB at large scales, setting the normalization scale as the pivot scale at $k_* = 0.05 \mathrm{Mpc}^{-1}$. The usual parametrization for the spectrum follows the evidence that at large scales the spectrum is almost scale-invariant. Thus
\begin{equation}
    \label{eq:Ps-1}
    \mathcal{P}_{\mathcal{R}}(k)=\mathcal{A}_s\left(\frac{k}{k_{*}}\right)^{n_{s}-1},
\end{equation}

\noindent with the CMB normalization dictating $\ln(10^{10}\mathcal{A}_s) = 3.044 \pm 0.014$ and the spectral index $n_s = 0.9649\pm 0.0042$ \citep{Planck:2018jri}.

Such prescription for the power spectrum would produce a tiny amplitude of fluctuations and produce an extremely small number of PBHs (see e.g.~\citep{Carr:1994ar,Emami_2018}). The spectrum, however, may suffer modifications at small scales, where features in the potential may arise and impact the amplitude of fluctuations significantly (see e.g.~the setting in Sec.~\ref{sec:example}). It is precisely such possibilities what we aim to explore, and possibly constrain, through the probability of PBH formation in a fertile scenario; the reheating epoch. 

\subsection{Inflaton evolution after inflation}\label{reheating}

\subsubsection{Reheating}
The reheating of the universe refers to the process through which the energy stored in the inflaton field, responsible for driving the inflationary expansion, is transferred to other particles present in the universe. This energy transfer takes place at the end of the inflationary period (see Fig.~\ref{fig:inflation}) and is believed to have created the necessary conditions for the formation of primordial nuclei and structures within the universe.\footnote{This is the simplest and most widely accepted mechanism by which it is assumed that the reheating of the universe took place. There are however,  alternative scenarios such as a reheating generated by moduli fields, a massive metastable particle, or reheating from PBH evaporation (see for example~\citep{Allahverdi:2020bys}). In this paper we will focus on the simplest case of reheating due to the inflaton field, although many of the results shown here can be easily extended to the other mechanisms.} Historically, reheating was first treated perturbatively \citep{Abbott:1982hn,PhysRevLett.48.1437} and in this section we review a simple version of this process. 

For transfering energy, it is usual to consider a non-minimal coupling of the inflaton with, say, a second scalar field $\chi$ through an interaction in the lagrangian. That is, 
\begin{equation}
    \mathcal{L}_{\rm int} = -g\Sigma \varphi\chi^2,
\end{equation}
where $g$ is a dimensionless coupling constant and $\Sigma$ is a mass term. The decay rate of the inflaton field into $\chi$ particles is thus given by {\citep{Greene:1998nh}}
\begin{equation}\label{eq:gam}
    \Gamma = \frac{g^2 \Sigma^2}{8\pi m},
\end{equation}
where $m$ is the ``effective" inflaton mass. The energy loss of the inflaton through its conversion to $\chi$ particles can be approximated by the following Klein-Gordon equation {\citep{doi:10.1146/annurev.nucl.012809.104511}}
\begin{equation}
    \ddot\varphi_b +3H\dot\varphi_b +\Gamma\dot\varphi_b = - dV(\varphi)/d\varphi.
\end{equation}

\begin{figure}[H]
    \centering
    \includegraphics[width=3.5in]{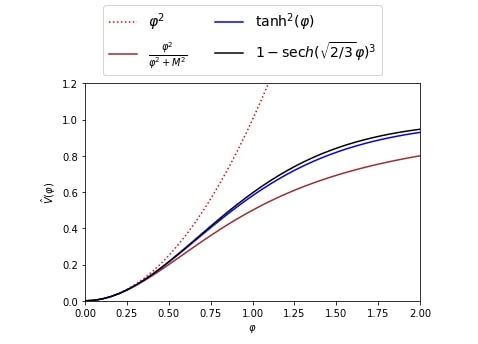}
    \caption{\footnotesize{Different normalized inflationary potentials as a function of $\varphi$. The dotted red line corresponds to the typical chaotic potential $V(\varphi)\sim \varphi^{2}$. The brown, blue, and black lines correspond to a polynomial \cite{Kallosh_2022, Iacconi:2023slv}, T-models \cite{JohnEllis_2013,Kallosh:2013yoa,RenataKallosh_2013}, and generalized $\alpha$-attractor T-models \cite{German2021, LinaresCedeno:2022rbq}, respectively. In the plot, we used $M = 1$.}}
    \label{fig:potentials}
\end{figure}

\noindent A typical form of the potential used for the reheating epoch is the quadratic potential, $V(\varphi) = {m^2}\varphi^2/{2}$, since, in order to have efficient reheating, it is necessary for the inflaton to oscillate around its global minimum. This precise form of the potential is not necessarily followed during the inflationary epoch and, in fact, for inflation to yield the correct observations given by Planck data \citep{Planck:2018jri}, a power law of the form $V(\varphi)\sim \varphi^\alpha$, with $\alpha<1$ is required. However, there are many realizations that meet this condition at large values of the field and that converge to the simple quadratic form at small field values. Some examples of such potential are shown in Fig.~\ref{fig:potentials}. This class of potentials are considered in this work as our main study cases.

To proceed with the analysis, let us note that in the case of an small coupling constant $(\Gamma\ll H)$, the interaction term $\Gamma$ can be neglected and the equation of motion of the inflaton reduced to
\begin{equation}  \ddot\varphi_b+3H\dot\varphi_b+m^2\varphi_b \simeq 0.
\end{equation}
This form suggests that in the limit in which $m\gg H$ (a limit that is fulfilled during the reheating epoch) the field $\varphi_b$ experiences damped oscillatory motion about $\varphi_b = 0$,
\begin{equation}\label{eq:phi}
    \varphi_b(t) = \sqrt{\frac{8}{3}}\frac{M_{\rm Pl}}{m}\frac{1}{t}\sin(mt).
\end{equation}
Here all quantities are displayed in units of the Planck mass $M_{\rm Pl}$.
In terms of the scale-factor and averaging over several oscillations we obtain
\begin{equation}\label{eq:be}
    \rho_b(a) = \rho_{\rm end}\left(\frac{a_{\rm end}}{a}\right)^3,
\end{equation}
where the background energy density is $\rho_b = \dot\varphi_b/2+V(\varphi_b)$, and we have used the subscript $_{\rm end}$ to refer to quantities evaluated at the end of inflation. The above result shows a pressure-less matter behavior, which in the background can be adopted while the inflaton dominates the overall energy density.

Once the Hubble expansion rate decreases to values comparable to $\Gamma$, the $\chi$-particle production becomes efficient, and the energy associated with the inflaton is transferred to the field $\chi$. The temperature at the time at which $\Gamma = H$ is known as the reheating temperature and is given by $T_R \sim \sqrt{\Gamma m_{\rm Pl}}$.

As shown in Eq.~\eqref{eq:gam}, $\Gamma$ is proportional to the square of $g$ and, since typically $g\ll 1$, we should expect that the reheating temperature occurs at low energy scales (low compared to the energy scale of inflation, at around $10^{14}~\rm{GeV}$). This temperature can be as low as the scale of Big Bang Nucleosynthesis (BBN, at around $10~\rm{MeV}$), since this is the maximum energy scale at which we have evidence of a radiation-dominated universe. This allows us to consider a scenario where reheating could have lasted a few $e$-foldings. This so called slow-reheating process is the scenario we explore in this review. We mainly look at the implications of a slow-reheating brings on the formation of PBHs, the conditions for collapse in this scenario, and the kind of inflationary models that can be constrained with this observable.

To start, it is important to define the stages of the evolution of overdensities of the inflaton field during slow-reheating. In the canonical mechanism, the $k$-modes associated with the quantum fluctuations of the inflaton field have a fixed amplitude when they stretch beyond the Hubble horizon during the accelerated expansion phase. Once inflation ends, these modes reenter the cosmological horizon after
\begin{equation}
    N_{\rm HC}(k) = 2\ln\left(\frac{k_{\rm end}}{k}\right)
\end{equation}
$e$-folds of expansion, where the subscript $_{\rm HC}$ refers to quantities evaluated at the horizon crossing time. Inside the cosmological horizon, two regimes are distinguished, which are separated by the scale $k_Q$, usually referred to as the quantum Jeans scale or simply the Jeans scale, given by \citep{PhysRevD.92.023510}
\begin{equation}
    k_Q = (16\pi G\rho_0 m^2a^4)^{1/4}.
\end{equation}

Inhomogeneities can be characterised by the density contrast $\delta= {\delta\rho}/{\rho_b}$, for which the time evolution (neglecting the decaying mode) is 
\begin{equation}\label{delta_a}
    \delta(a;k) = \delta_{\rm HC}(k)\frac{a}{a_{\rm HC}}, \ \ \ \text{for} \ \ k_{\rm H}<k<k_Q.
\end{equation}

\noindent Here $k_{\rm H} = aH$ is the scale associated to the size of the cosmological horizon. Thus, density fluctuations with a characteristic scale $k > k_Q$ undergo damping via oscillations, while at scales $k_H<k < k_Q$  fluctuations experience a growth in amplitude proportional to the scale factor\footnote{This resonance band appears naturally when doing a perturbative analysis of the system, where the resonance is obtained from a Mathieu-type equation \citep{Kofman:1997yn,Jedamzik:2010dq}.}.

\subsubsection{Preheating}

The preheating instability is a non-perturbative mechanism that arises in theories with non-minimal couplings between the inflaton and other fields, say $\chi$ \citep{ElBourakadi:2021blc}. The dynamics of the inflaton during the preheating process can be described by the Mathieu equation, {which is related to periodic or quasi-periodic oscillating systems} \citep{Bassett:2005xm, Kofman:1994rk, McLachlanN.W.NormanWilliam1964Taao}. Furthermore, its solutions exhibit exponential instabilities, that is, $\chi_{k} \propto \exp(\mu_{k}^{(n)}mt)$ within a series of resonance bands, located at specific frequency ranges $\Delta k^{(n)}$ (here labeled by the integer index $n$). Such instabilities lead to an exponential growth in the occupation numbers of quantum fluctuation modes, denoted as $n_{\bar{k}}(t) \propto \exp(2\mu^{(n)}_{k}mt)$, which are interpreted as the production of $\chi$ particles \citep{Kofman:1997yn, Bassett:2005xm, Abbott:1982hn, Lyth:2009zz}. In short, preheating describes the process through which the energy density of the created particles, calculated within the above formalism, is extracted from the energy density of the oscillating inflaton field.

\section{PBH formation during preheating}\label{sec:preheating}

%In order to present a brief description about the origins of PBHs formation during preheating, it is necessary to discuss more about the theory of the transition from inflation to thermalization. For years this has been investigated with fully non-linear numerical lattice simulations plus different techniques of classical field dynamics \citep{Kofman:1994rk, Felder:2000hq, Micha:2004bv, Khlebnikov:1996mc, Khlebnikov:1998sz, Felder:2000hr, Dux:2022kuk, Repond:2016sol}. In last years preheating has been the escenario to use lattice field theory simulations that evolve the scalar field equations on a homogeneous background, and neglect the backreaction of inhomogeneities on the local spacetime metric. These effects become important when overdensities grow large enough for gravity to be of the same order as self interactions in the field \citep{Figueroa:2020rrl, Figueroa:2021yhd, Frolov:2008hy}. This innovative production mechanism has the potential to generate a significant density of PBHs, which could potentially explain all or a portion of dark matter. 

The production of PBHs during preheating was first studied in Ref.~\citep{Green:2000he}. In such work, the authors studied a two-field chaotic inflation model and found that for a wide range of parameters the resonant amplification of modes during preheating leads to an overproduction of PBHs, before backreaction  terminates the resonance. 

In order to handle the non-perturbative {and} non-minimal interaction between fields, the Preheating process has been modeled through lattice field theory simulations that evolve the scalar field equations on a homogeneous background \citep{Kofman:1994rk, Khlebnikov:1996mc, Khlebnikov:1998sz, Felder:2000hq, Felder:2000hr, Micha:2004bv,    Repond:2016sol, Dux:2022kuk}, though neglecting the backreaction of inhomogeneities on the local spacetime metric\footnote{Backreaction effects become important when overdensities grow large enough for gravity to be of the same order as self interactions in the field \citep{Figueroa:2020rrl, Figueroa:2021yhd, Frolov:2008hy}.}. 

With the use of lattice simulations came the development of computational techniques that captured the non-linear aspects of the problem and showed the importance of various factors on PBH formation \citep{Shibata:1999zs, Niemeyer:1999ak,  Harada:2013epa, Nakama:2013ica, Nakama:2014bxa}. These factors encompassed the equation of state, the nature of the inflationary potential, and the presence of additional fields or interactions\footnote{In fact, in Refs.~\citep{1978SvA....22..129N, 1980SvA....24..147N}, this dependence was investigated, illustrating within a spherically symmetric {analysis} the process of PBHs formation and their accretion for different equation state values.}. Importantly, even small inflaton self-interactions can accumulate over multiple oscillations, triggering the resonant growth of non-zero momentum inflaton modes, which constantly interact with the homogeneous component. This process of self-fragmentation in the inflaton field not only redistributes the initial energy density but also leads to the formation of localized soliton-like structures referred to as oscillons (see e.g. \citep{Bogolyubsky:1976yu, Gleiser:1993pt, Linde:1990flp,Kolb:1993hw, Copeland:1995fq, Honda:2001xg, Fodor:2006zs, Hong:2017ooe, Amin:2011hj, Antusch:2017flz}). The presence of large numbers of oscillons has motivated the search for PBH production and the implications for reheating, placing constraints on various single-field inflation models and other models accommodating oscillon solutions (e.g. \citep{Cotner:2018vug, Cotner:2019ykd}). Specifically, Ref.~\citep{Cotner:2018vug} shows that the fragmentation of the inflaton into oscillons can give rise to the formation of PBHs in single-field inflation models or other models permitting oscillon solutions. Subsequently, Ref.~\citep{Suyama:2004mz} showed that PBH production in preheating does not exceed astrophysical bounds because the mass of PBHs is small enough to evaporate before BBN. As a result, these PBHs are not constrained by observation, even if they are overproduced, unless they leave behind Planck mass relics  \citep{PhysRevD.46.645}.

It has also been argued that the assumption of a Gaussian probability distribution for density perturbations at horizon crossing is crucial. Since the density perturbations that lead to PBH formation are very rare and sensitive to the tail of the distribution, on average, PBHs are not overproduced during the violent non-equilibrium phase of preheating that follows the end of many inflationary models. In Ref.~\citep{Green:2000he} a linear approximation estimated the time when backreaction becomes significant and when the amplitude of density perturbations surpasses a certain threshold, treating them separately. This leads to a criterion for PBH formation. It is important to note that even a small error in determining the backreaction time can lead to incorrect conclusions due to the exponential growth of perturbation amplitude. This discrepancy highlights the sensitivity of the results to the precise timing of backreaction and its potential impact on the predictions for PBH production or overproduction. On the other hand, Ref.~\citep{Torres-Lomas:2013uzl} modified a version of HLattice to numerically solve the relevant equations of motion and analyze the mass variance as a means to explore the formation of structures during the preheating phase. The study revealed that preheating has the potential to generically produce PBHs. However, the results highlighted the influence of the smoothing scale values and emphasized the need for backreaction, to confirm the obtained results. Subsequently, Ref.~\citep{Caravano:2022epk} found strong backreaction effects in the system, invalidating the standard perturbation theory approach. They also observed that the non-Gaussianity of the comoving curvature perturbation is large in the linear regime but gets suppressed as the dynamics become nonlinear. This suppression of non-Gaussianity relaxes the bounds on PBH overproduction, allowing instead for an observable gravitational wave signal at interferometer scales.

\section{PBH formation during slow-reheating: Direct gravitational collapse}\label{sec4}

The transition of the universe to the standard Big Bang cosmology prior to the BBN epoch can be achieved through a variety of mechanisms. One possibility is the fragmentation of the inflaton condensate into its own quanta, triggered by self-resonance~\citep{Amin:2010xe, MustafaA.Amin_2010, PhysRevD.97.023533, Fukunaga:2019unq}. In the same context, as mentioned above, particles coupled to the inflaton can be resonantly produced~\citep{Kofman:1997yn,Kofman:1994rk} leading to prompt thermalization \citep{PhysRevLett.119.061301} or a potential oscillon dominated epoch \citep{MustafaA.Amin_2010,Amin:2010xe,PhysRevLett.108.241302,PhysRevD.97.023533}. During this epoch, the formation of PBHs is possible due to the gravitational collapse of the perturbations that were resonantly amplified. 

In contrast, if parametric resonance is not present, the particle production occurs through a more gradual, perturbative processes (as described in e.g. \citep{Abbott:1982hn,PhysRevLett.48.1437}). As discussed in Section \ref{reheating}, these processes can take place during a relatively long period of expansion when the universe is governed by a nearly homogeneous condensate, in a (nearly) $ \varphi^2$ potential, and when $\Gamma\ll H$. Eventually, the condensate fragments due to the gravitational growth of perturbations \citep{KarstenJedamzik_2010,RichardEasther_2011}, in what is dubbed a primordial structure formation process, where inhomogeneities virialize.
When the matter concentration is sufficiently high, the hoop conjecture prescribes that some of these structures may instead collapse and form PBHs. The mass of such black holes should be close to the mass of the cosmological horizon evaluated at the time of horizon crossing, conveniently expressed as:
\begin{equation}\label{mass}
\frac{M_{\rm PBH}(k)}{7.1\times 10^{-2}\rm{g}} = \gamma\frac{1.8\times 10^{15}\rm{GeV}}{H_{\rm end}}\left(\frac{k_{\rm end}}{k}\right)^3,%(1+\delta_{\rm end}(k)),
\end{equation}

\noindent Here, $\gamma$ is a constant that encrypts the efficiency of the collapse. The precise value of $\gamma$ ought to be determined through numerical calculations currently in progress for the reheating scenario (partial progress has been reported in \citep{deJong:2021bbo,Padilla:2021uof}). For the sake of the argument, we will adopt the value $\gamma = 1$. In this and the following two sections, we review the conditions for which PBH formation may occur in a slow-reheating scenario.

Since the inflaton during reheating behaves like pressureless matter, one may be tempted to extrapolate the perfect fluid criterion $\lim_{w \to 0} \delta_{\rm th} \to w$\citep{Harada:2013epa} and deduce a copious formation of black holes, following the Press-Schechter formalism usually employed in the standard caculation, and as exemplified for this case in Appendix A.  Such approach is an over-simplification of the problem since, as we have mentioned in Sec.~\ref{sec:scales}-- and will discuss in more depth in the following, the inflaton presents an effective pressure due to its quantum nature, which prevents the formation of PBHs from overdensities of infimum amplitude. Moreover, the collapse criteria used for dust-like overdensities can be adopted and complement those derived for a cosmological scalar field. In the following we review the diverse criteria with the aim of assessing PBH formation in a slow-reheating scenario. 

\subsection{The sphericity criterion}\label{sphericity}

One of the most widely used criteria to describe the formation of PBHs in a slow-reheating scenario was discussed early in the develpment of the theory \citep{Polnarev:1985btg,Khlopov:1980mg}. Physically, this criterion limits the configurations of initial pressureless overdensities to be sufficiently close to spherical symmetry, so as to collapse onto a black hole.
Inspired by this, Ref.~\citep{Harada:2016mhb} presents a more detailed analysis of the so-called sphericity criterion for the collapse of overdensities. The latter article investigates the formation of PBHs in the matter-dominated phase of the Universe, where nonspherical effects in gravitational collapse play a crucial role. The authors apply the Zel'dovich approximation \citep{1970A&A.....5...84Z}, Thorne's hoop conjecture \citep{wheeler1972magic}, and Doroshkevich's probability distribution \citep{osti_4475244} to derive the production probability $\beta_0$ of PBHs. 
In summary, in the limit of a small variance of $\delta$ evaluated at the horizon crossing time, the relation
\begin{equation}\label{sigma1}
    \beta_0\simeq 0.05556\sigma^5,
\end{equation}

\noindent approximates the initial abundance of PBHs as a function of the variance $\sigma^2$ in a dust dominated universe. Note that this is not directly linked to a threshold amplitude, but instead this prescription alone may overproduce PBHs even for a scale-invariant power spectrum, if reheating lasts long enough \cite{Hidalgo:2017dfp}.
%\subsection{Juan Carlos \textit{et. al.} criterion}

\subsection{Conservation of angular momentum criterion}\label{sec:angular_momentum}

The initial angular momentum of overdensities plays an important role in the formation of PBHs. In Ref.~\citep{Harada:2017fjm} it was shown that this can lead to a significant suppression of the production rate. In particular, it was found that the limit on the ammount of angular momentum allowed for collapse provides a threashold value for PBH formation which complements  the sphericity criterion. Specifically, the production probability $\beta_0$ is restricted to
\begin{equation}\label{sigma3}
    \beta_0 \simeq 1.9\times 10^{-7}f_q(q_c)\mathcal{I}^6\sigma^2\exp\left[-0.15\frac{\mathcal{I}^{4/3}}{\sigma^{2/3}}\right],
\end{equation}
where $\mathcal{I}$ is a parameter of order $O(1)$ and $f_q(q_c)$ is the fraction of mass with a level of quadrupolar asphericity $q$ smaller than a threashold $q_c$. Comparing this result with \eqref{sigma1}  and assuming $\mathcal{I} = f_q(q_c) = 1$ (as assumed in Ref.~\citep{Harada:2017fjm}) it is evident that the angular momentum criterion of Eq.~\eqref{sigma3} is more stringent than the sphericity criterion of Eq.~\eqref{sigma1} for a standard deviation $\sigma \lesssim 0.005$ (the relevance and matching of these constraints is illustrated in Fig.~\ref{fig:beta_sigma}).

We conclude this section by mentioning that according to the study conducted in Ref.~\citep{deJong:2023gsx} on the formation of spinning PBHs during an early matter-dominated era, the efficiency of mass transfer was found to be approximately 10\%, while the efficiency of angular momentum transfer was estimated to be around 5\%. That reference further suggests that unless the matter era is short, the final dimensionless spin of PBHs is expected to be negligible. 

\subsection{Reheating time criterion}\label{sec:rtc}

Another employed criterion for characterizing the generation of PBHs arising from an slow-reheating epoch can be found in Ref.~\citep{Martin_2020} (see also, e.g.~\citep{ Carr:2018nkm,Goncalves:2000nz}, for earlier work that considers this criterion of PBH formation). Unlike the previous criteria that consider the morphology and angular momentum of the initial inhomogeneity, this criterion is mainly focused on studying the time required for the collapse of  configurations. In this way, one can impose a condition on the contrast density evaluated at the horizon crossing time that a perturbation must meet in order to form a PBH. In this section, we will review the most important results of such work, considering that the reader interested in the details of the calculations presented here will be able to review the original papers.

%As discussed previously in section II, after inflation ends, the inflaton enters a phase of oscillation at the bottom of its quadratic potential. During this epoch, the evolution of perturbations depends on the scales considered.
%On small scales, the situation can be significantly distinct. This was demonstrated by Karsten J. \textit{et al.} in \citep{Jedamzik:2010dq}, where it was shown that for modes that fulfill the following inequality 
%\begin{equation}
    %aH < k < a\sqrt{3Hm}\,,
    %\label{eq: band_inequal}
%\end{equation}
%where this corresponds to being within the first instability band of Mathieu equation. The instability arises when the physical wavelength of a mode is smaller than the Hubble radius during reheating, but larger than a new scale defined by $\sqrt{3Hm}$.

%Once within the instability band, the fluctuations get strongly amplified, such that the density contrast grows linearly with the scale factor. Effectively, they thus behave as pressureless matter perturbations in a pressureless matter universe. during this epoch, cosmological perturbations at the amplified scales may collapse into PBHs. This epoch is referred to as the “instability phase”. During this epoch, cosmological perturbations at the amplified scales may collapse into PBHs. When the inflaton decays into other degrees of freedom (or when the PBHs take the inflaton over, see below), the instability stops, and the density of black holes evolves under various physical effects (cosmic expansion, Hawking evaporation, accretion, merging, etc.).

As previously argued in Sec. \ref{reheating}, modes within the resonance band $k_H<k<k_Q$ are expected to behave as standard pressureless matter fluctuations, which may therefore collapse to form a primordial structure such as a PBH. The time {$t_c$} required for such collapse {in the spherical collapse model} is given by \citep{Goncalves:2000nz}:
\begin{equation}
\Delta t_{\rm c}(k)\equiv t_c(k)-t_{\rm bc}(k) = \frac{\pi}{H_{\rm bc}(k)\delta_{\rm bc}(k)^{3/2}}\,.
    \label{eq:time_to_form}
\end{equation}
Here the subindex $_{\rm bc}$ is related to quantities evaluated at the time when the mode $k$ transitions across the instability band. Note that there are two ways in which a $k-$mode may enter the instability band. First are those scales which exited the cosmological horizon during inflation and reenter (entering from above) after the inflationary epoch. The second possibility is for those scales that enter the instability band from below, due to the fact that the quantum Jeans scale decreases in size as $\sim a^{-1/4}$. Following the work of Ref.~\citep{Martin_2020}, in this section we shall only concentrate on the scales which enter the instability band from above and thus we equate $t_{\rm bc} = t_{\rm HC}$. Additionally, we can also re-express Eq.~\eqref{eq:time_to_form} in terms of the number of $e$-foldings. Considering that during a matter-dominated universe we have $H = 2/(3 t)$ and assuming $3\pi/(2\delta_{\rm HC}^{3/2}(k))\gg 1$, which is a good approximation in the perturbative regime, we obtain
\begin{equation}\label{eq:N_c}
   \Delta N_c(k) \equiv N_c(k)-N_{\rm HC}(k)\simeq \ln(2.81\delta_{\rm HC}^{-1}(k))
\end{equation}

We can identify the last mode to enter into the instability band from above as $k_\Gamma = a_{\Gamma}H_{\Gamma}$, where subindex $_\Gamma$ indicates quantities evaluated at the time the inflaton field decays. We can then relate the scale $k_\Gamma$ with the scale at the end of inflation, $k_{\rm end}$ by $k_\Gamma/k_{\rm end} = (\rho_{\Gamma}/\rho_{\rm end})^{1/6}$. Consequently, the scales that can collapse gravitationally during the slow-reheating epoch to form a PBH are those within the interval
\begin{equation}
    \left(\frac{\rho_{\Gamma}}{\rho_{\rm end}}\right)^{1/6} < \frac{k}{k_{\rm end}} < 1\,.
\end{equation}

The condition adopted in \citep{Martin_2020} to determine if a scale $k$ can collapse to form a PBH was to require that the time for the collapse of the perturbation be smaller than the total period that the slow-reheating epoch spans. Noting that such a time can be expressed as
\begin{equation}
    t_{\Gamma} - t_{\rm HC}(k) = \frac{2}{3H_{\rm HC}(k)}\left[\left(\frac{a_{\Gamma}}{a_{\rm HC}(k)}\right)^{3/2} - 1\right]\,,
\end{equation}
and using Eq.~\eqref{eq:time_to_form}, we obtain that the condition for a perturbation to collapse is given by
\begin{equation}
\delta_{\rm th}(k) < \delta_{\rm HC}(k) < 1\,,
\label{eq: condition}
\end{equation}
where
\begin{equation}
    \delta_{\rm th}(k) \equiv \left(\frac{3\pi}{2}\right)^{2/3}\left[\left(\frac{k}{k_{\rm end}}\right)^{3}\sqrt{\frac{\rho_{\rm end}}{\rho_{\Gamma}}}-1\right]^{-2/3},
    \label{eq:delta_th}
\end{equation}
and the maximum value in this context is determined by the requirement for the horizon-crossing fluctuation to be within the perturbative regime\citep{Martin_2020}, imposing this upper bound should not significantly impact the abundance of PBHs, considering that the amplitudes of density contrasts are exponentially suppressed, preventing an overly abundant production ~\citep{Harada:2017fjm,Niemeyer:1997mt}.

%\JCH{Is this bound justified in \citep{Carr:2018nkm}? Why "in relation to abundances of PBHs"?}\TDG{You're right! I made the correction the cite is the next:.~\citep{Martin_2020}, the relation to abundance is just that as we have seen, the density perturbation $\delta$ has a finite maximum and one minimum value that is determined by the Eq. \eqref{eq:delta_th}, the condition $\delta_{\text{th}}(k) < \delta_{\text{HC}}(k) < 1$ effectively acts as a control mechanism for the abundance of PBHs. It ensures that the perturbations leading to PBH formation are within a specified range, preventing an overly abundant production of PBHs when you do use the Press–Schechter formalism~\citep{Harada:2017fjm,Niemeyer:1997mt}, I guess that this conservative term refers to indicating a careful and restricted approach when setting the conditions for the PBHs formation, focusing on avoiding an overestimation of the abundance of PBH by considering a well-defined subset of modes that enter the perturbative regime. But I will delete this to prevent confuse.}

To calculate the abundance of PBHs $\beta_0$ formed in this context, we can adopt the Press-Schechter formalism (see Appendix \ref{AP:PS} for details):
\begin{equation}\label{eq:ps}
    \beta_0(M_{\rm PBH}) = -2M_{\rm PBH}\frac{\partial R}{\partial M_{\rm PBH}}\frac{\partial \mathbb{P}[\delta>\delta_{\rm th}]}{\partial R},
\end{equation}
where $\mathbb{P}[\delta >\delta_{\rm th}]$ is the probability that a smoothed density field exceeds the threshold value $\delta_{\rm th}$ and is given by
\begin{equation}\label{Pdelta2_main}
    \mathbb{P}[\delta>\delta_{\rm th}] = \frac{1}{2}\text{erfc}\left(\frac{\delta_{\rm th}}{\sqrt{2}\sigma(R)}\right).
\end{equation}
In the above expression $R = 1/k$, $\erfc(x)=1-\erf(x)$ is the complementary error function, and $\sigma(R)$ is, as previously defined, the variance of $\delta$ evaluated at the horizon crossing time. 

It is important to note that this formation criterion assumes that all perturbations with a collapse time shorter than the remaining time for reheating will gravitationally collapse to form PBHs. Consequently, this criterion is likely to overestimate the actual abundance of PBHs formed, $\beta_0$. This is due to the omission of significant physical effects discussed throughout this review. In the follwing we focus on the physics of fluctuations from an oscillating scalar field, while in Sec.~\ref{Sec:Slow-reheat-phbs} we address the PBH formation criteria in such environment.

\section{Dynamics and structure formation in a slow-reheating epoch}

    In the preceding section, we described the process of PBH formation during a period of slow-reheating, assuming a dust-like behavior for the governing inflaton field. However, such approximation oversimplifies the dynamics of a cosmological scalar field, as it neglects its inherent quantum nature. In order to describe more accurately the evolution of the post-inflationary epoch one must solve the Einstein-Klein-Gordon (EKG) system of equations in a cosmological background. However, such task presents the practical difficulties described below.
    
    The post-inflationary regime for an (almost) free field is subject to the condition $m\gtrsim H_{\rm end}$. As indicated by Eq.~\eqref{eq:phi}, the  oscillation frequency of the homogeneous inflaton field is precisely $m$. On the other hand, the universe evolution is characterized by the Hubble time $1/H \sim a^{3/2}$. Thus, a few $e$-folds after the end of inflation, the condition turns into $1/H\gg 1/m$. This thus stipulates two dissimilar characteristic time scales in the numerical evolution of the complete EKG equations, which turn the evolution over several Hubble times computationally unfeasible. Efforts in this direction are found in Refs.~\citep{Alcubierre:2015ipa,Rekier:2015isa,deJong:2021bbo}.

    Here we follow a different approach, acknowledging the similarities between a slow-reheating epoch and the phase of structure formation in the scalar field dark matter (SFDM) model. Such parallelism was initially proposed in \citep{Musoke:2019ima} and has extended in subsequent studies \citep{Niemeyer:2019gab, Eggemeier:2020zeg, Eggemeier:2021smj, Eggemeier:2022gyo, Chavanis:2022fvh, DeLuca:2021pls, Padilla:2021zgm, Hidalgo:2022yed, Padilla:2023lbv, Carrion:2021yeh, Hidalgo:2017dfp}. Taking advantage of the extensive literature on SFDM models, we aim for a better understanding of the universe right after inflation. In the upcoming sections, we will review various SFDM results that can be adapted to the slow-reheating scenario. With such results at hand, in the following section, we formulate the conditions under which the primordial structures of this period may undergo gravitational collapse, resulting in the formation of PBHs.

%We can utilize certain approximations to simplify the treatment of the system. For instance, when a particular scale becomes non-linear, we should expect it to be found within subhorizon scales, its bulk motion exhibit a nonrelativistic behavior, and its occupation numbers become high. These specific circumstances create an excellent opportunity to apply the (Newtonian) Schrödinger-Poisson (SP) formalism. In this formalism, matter is represented by the non-relativistic wavefunction $\psi$, and the gravitational potential $V_N$ is determined by solving the Poisson equation. 

\subsection{The Schrodinger-Poisson picture}

Let us look at a few approximations to simplify the treatment of the post-inflationary universe. For instance, when a particular scale becomes non-linear, it is expected to be well within the Hubbet horizon and with a bulk motion exhibiting a nonrelativistic behavior, with large occupation numbers. These specific characteristics in the system are the precise requirements to apply the Newtonian approximation, where the EKG system can be reduced to the Schrödinger-Poisson (SP) system of equations. In such approximation, matter is represented by the non-relativistic wavefunction $\psi$, and the (Newtonian) gravitational potential $\Psi$ is determined by solving the Poisson equation. Below we quickly review how the Newtonian description of the system is reached.

In the slow-reheating era, it is possible to describe gravity through the weak-field approximation. Well within the cosmological horizon, we adopt a spatially flat background metric and deal with scalar perturbations in the Newtonian gauge (see e.g.~Ref.~\citep{Malik:2008im}),
\begin{equation}
    g_{00} = -(1+2\Psi(\text{\textbf{x}},t)), \ \ \ g_{0j} = 0,
    \nonumber
    \end{equation}
    \begin{equation}
    g_{ij} = a\delta_{ij}(1+2\Phi(\text{\textbf{x}},t)).
\end{equation}
Considering that the anisotropic stress of a minimally coupled scalar field vanishes, we can identify the Newtonian potential as $\Psi = -\Phi$.  This allow us to write the Einstein-Hilbert action for subhorizon scales ($k\gg aH$) and by considering quantities at first order in the potential and at second order in spatial derivatives as \citep{Niemeyer:2019aqm}
\begin{eqnarray}
    S_{\rm EH} =&& \int dx^4a^3\left[-\frac{(\partial_i \Psi)^2}{8\pi G a^2}+\left(\frac{1}{2}(1-4\Psi)\dot\varphi^2-\frac{1}{2a^2}{(\partial_i\varphi)^2}\nonumber\right.\right. \\
    &&
    \left.\left.-(1-2\Psi)\frac{m^2}{2\hbar^2}\varphi^2\right)\right] \label{eq:eh}
\end{eqnarray}
%If we now rearrange the above action by considering quantities at first order in the potential and at second order in spatial derivatives, we obtain \citep{Niemeyer:2019aqm}
%\begin{eqnarray}
    %S_\varphi =&& \int dx^4a^3\left[\frac{1}{2}(1-4\Psi)\dot\varphi-\frac{1}{2a^2}(\partial_i\varphi)^2\right.\nonumber\\
    %&&-\left.(1-2\Psi)\frac{m^2}{2\hbar^2}\varphi^2\right]\label{eq:eh}
%\end{eqnarray}
Further simplifcation is achieved through the following considerations; while $\varphi$ oscillates at a frequency $m$, the density field changes slowly within the nonrelativistic regime. To account for the rapid oscillations, we can introduce the complex field $\psi$,  as follows:
\begin{equation}
    \varphi = \frac{\hbar}{\sqrt{2ma^3}}(\psi e^{-imt/\hbar}+\psi^* e^{imt/\hbar}).
\end{equation}
With such definition, disregarding oscillatory terms that involve powers of $\exp(\pm imt/\hbar)$, and subsequently incorporating the simplifying assumptions above justified, namely $\dot\psi\ll m\psi$ and $m\gg H$, Eq.~\eqref{eq:eh} is simplified to
\begin{eqnarray}
    S = \int d^4x\left[\frac{i\hbar }{2}(\dot\psi\psi^*-\psi\dot\psi^*)-\frac{\hbar^2(\partial_i\psi)(\partial_i\psi^*)}{2ma^2}\nonumber \right.\\
    \left.-m(\psi\psi^*-\langle\psi^*\psi\rangle)\Psi-\frac{a}{8\pi G}(\partial_i \Psi)^2\right].
\end{eqnarray}
where we write explicitly the $\hbar$ factors to restore units and emphasize the quantum nature of the system here described.

After varying the above action $S$ with respect to the gravitational potential $\Psi$ and the field $\psi$, we finally arrive at the SP system of equations:
\begin{subequations}\label{eq:sp}
    \begin{equation}
        i\hbar\partial_t \psi = -\frac{\hbar^2}{2ma^2}\nabla^2\psi +m\Psi\psi,
    \end{equation}
    \begin{equation}
        \nabla^2\Psi = \frac{4\pi G}{a}(\rho-\langle\rho\rangle).
    \end{equation}
\end{subequations}
Here $\langle\rho\rangle$ is the smooth background value of the density of the scalar field, $\langle\rho\rangle = m\langle\psi\psi^*\rangle$.  

\subsection{Quantum hydrodynamics equations}

One of the great advances in the study of the SP system was the realisation that this pair of equations can be reformulated like classical hydrodynamics. The hydrodynamic version of the SP system introduces an additional quantity $Q$, known as the ``quantum potential" or, in some instances, the ``Bohm potential" \citep{PhysRev.85.166,PhysRev.85.180}. As a result, these equations are commonly named as the ``quantum hydrodynamics" (QHD) equations (we recommend \citep{wyatt2005quantum} for a textbook presentation of these equations). It is important to emphasize that both the hydrodynamic and the field expressions of the SP system, are equivalent and offer robust methods for addressing the nonlinear dynamics of a cosmological scalar field. Let us outline in this section two methods for deriving these QHD equations.

\subsubsection{The Madelung-Bohm formulation of quantum hydrodynamics}

We can consider a Madelung transformation \citep{Madelung1927QuantentheorieIH} of the form
\begin{equation}
    \psi = \sqrt{\frac{\rho}{m}}e^{im\theta/\hbar} = \sqrt{n}e^{im\theta/\hbar}.
\end{equation}
If we define the bulk flow velocity of the field \textbf{v} as \textbf{v}=$\nabla\theta$, we can reexpress the SP system of equations as
\begin{subequations}\label{eq:qhd}
\begin{equation}
\label{cons_ener}
\partial_t\rho+\frac{1}{a^2}\nabla(\rho\text{\textbf{v}}) = 0,    
\end{equation}
\begin{equation}
\label{cons_mom}
    \partial_t \textbf{v}+\frac{1}{a^2}(\textbf{v}\nabla)\textbf{v} +\nabla \Psi+\nabla Q = 0,
\end{equation}
where $Q$ is given by  
\begin{equation}
\label{Eq:Qdef1}
    Q = -\frac{\hbar^2}{2m^2a^2}\left(\frac{\nabla^2\sqrt{\rho}}{\sqrt{\rho}}\right)
\end{equation}
\end{subequations}
This set of equations constitute the QHD system.
The first of these equations, Eq.~\eqref{cons_ener}, corresponds to a continuity equation, typically found in classical fluid dynamics. Such an equation is used to describe the conservation of mass within the system. Additionally, the second equation, Eq.~\eqref{cons_mom}, is an Euler-like equation stating momentum conservation. However, instead of the conventional terms associated with a fluid pressure gradient, in these QHD equations, we encounter a novel potential $Q$ that encapsulates the quantum properties of the scalar field.

\subsubsection{Phase space formulation}\label{sec:PSF}

We can also obtain the QHD equations by taking {momentum} moments of the SP system of equations \citep{10.1143/PTP.11.341}. Following Ref.~\citep{Dawoodbhoy:2021beb}, we sketch the steps in the following. 

We define the Wigner function \citep{PhysRev.40.749} as
\begin{equation}\label{wigner}
    W(\textbf{x},\textbf{p},t) = \frac{1}{(2\pi\hbar)^3}\int \psi^*(\textbf{x}+\textbf{y}/2,t)\psi(\textbf{x}-\textbf{y}/2,t)e^{i\textbf{p}\cdot\textbf{y}/\hbar}d^3\textbf{y}.
\end{equation}
From this expression, the number density at a point in coordinate space is determined by the integral of $W$ over momentum space, 
\begin{equation}\label{n_wigner}
    {n(\textbf{x},t)} = \int W(\textbf{x},\textbf{p},t)d^3\textbf{p}.
\end{equation}

Moreover, the local average of a quantity $A$ over momentum space is computed through
\begin{equation}
    \langle A\rangle (\textbf{x},t) = \frac{1}{n(\textbf{x},t)}\int AW(\textbf{x},\textbf{p},t)d^3\textbf{p}.
\end{equation}
As an example, if we use the bulk velocity $\textbf{v } = \textbf{p}/m$ and we perform the above integration, we arrive at the same bulk velocity in terms of $\nabla\theta$ as before. One can also calculate the velocity dispersion tensor as follows
\begin{eqnarray}
    \sigma_{ij}^2(\textbf{x},t) &=& \frac{1}{n(\textbf{x},t)}\int \frac{(p_i-\langle p_i\rangle)(p_j-\langle p_j\rangle}{m^2}W(\textbf{x},\textbf{p},t)d^3\textbf{p}\nonumber \\
    &=&(\langle p_ip_j \rangle-\langle p_i\rangle\langle p_j\rangle)/m^2.
\end{eqnarray}
To derive the equation of motion for the Wigner function, we compute the partial time derivative ($\partial W/\partial t$) and incorporate it into the SP system of equations. The outcome is known as the Wigner-Moyal equation, which bears resemblance to the collisionless Boltzmann equation (CBE), also known as the Vlasov equation. For this particular case, the Wigner-Moyal equation introduces additional terms that encode the quantum characteristics of the scalar field.

When computing {momentum} moments of the Wigner-Moyal equation, it is possible to derive, from the $0^{\rm th}$ moment  the continuity equation, Eq. \eqref{cons_ener}, and from the $1^{\rm st}$ moment an Euler-like equation from classical fluid mechanics. Once again, in this formalism a new pressure-like term emerges, referred to as the ``quantum pressure" tensor, $\Pi_{ij}$, which is linked to the velocity dispersion tensor as
\begin{equation}\label{pp}
    \Pi_{ij} = \rho\sigma_{ij}^2 = \left(\frac{\hbar}{2m}\right)^2\left(\frac{1}{\rho}\frac{\partial\rho}{\partial x_i}\frac{\partial\rho}{\partial x_j}-\frac{\partial^2\rho}{\partial x_i\partial x_j}\right).
\end{equation}
It is worth mentioning that, in general, there is not a real difference between the $1^{\rm st}$ moment equation and Eq.~\eqref{cons_mom}, since both equations coincide once the quantum potential $Q$ is defined as
\begin{equation}\label{qp_pp}
    \frac{\partial Q}{\partial x_i} = \frac{1}{\rho}\frac{\partial \Pi_{ij}}{\partial x_j},
\end{equation}

\noindent which reduces to the explicit form of Eq.~\eqref{Eq:Qdef1}.

Note that Eqs.~\eqref{pp} and \eqref{qp_pp} stipulate that the force originating from the quantum potential term in the momentum equation 
 corresponds to the effective ``pressure" present in the momentum flux density associated with the internal distribution of momentum in the phase space derivation. In both situations, these ``quantum" terms arise from the kinetic term of the SP equations and represent the wave-like behavior of the scalar field. In practice this represents a force opposing gravitational collapse.% on a scale comparable to the de Broglie wavelength, $\lambda_{\rm dB}\equiv \hbar/(mv)$.
 
\subsection{Some important scales}\label{sec:scales}

As shown above, the hydrodynamic formulation of the SP equations is a system that closely resembles the description of the nonlinear dynamics of a pressure-less fluid, with the exception of an additional term that accounts for the quantum properties of the scalar field. 
%This prompts us to address the following inquiries:
%\begin{enumerate}
%    \item At what characteristic scale does the wave dynamics of the scalar field begin to exert a significant influence?
%    \item What is the characteristic timescale at which we should expect deviation from the simplified dust-like behavior?
%\end{enumerate}
In this formulation it is easy to see that the quantum potential term becomes important only when its contribution is comparable to the kinetic and gravitational potential. This happens for scales that fulfill $R\sim \lambda_{\rm dB}$, where $R$ ($\lambda_{\rm dB}\equiv \hbar/(mv)$) is the characteristic scale (de Broglie wavelength) of the configuration (see for example \citep{Niemeyer:2019aqm}). At scales $R\gg \lambda_{\rm dB}$ the expected behavior of the scalar field configurations should be very similar to that of a dust-like component, while for scales with $R\sim \lambda_{\rm dB}$  the quantum contribution must be considered.

In order to estimate the timescales at which the quantum characteristics of the scalar field become significant, one may look at the gravitational scattering time for wave scattering within a condensate\footnote{{A Bose-Einstein condensate is an exotic state of matter where particles clump together and behave as a single quantum entity. The idea in this studied context is that the inflaton must condense during the post-inflationary universe, forming structures (such as solitons) as a result of this process.}}. 
In the absence of external influences, the scattering rate $\Gamma_s$, inversely proportional to the time interval $\tau$, is dependent on the scattering cross section $\sigma_g$, the average relative velocity $\langle v\rangle = \sqrt{2}v$, and the number density $n=\rho/m$. Namely,  $\Gamma_s\propto \sigma_g\langle v\rangle n$. However, when the final state experiences macroscopic occupation, the rate is further enhanced by the scalar field phase space density, often referred to as the occupation number $\mathcal{N}$. Such enhancement is due to the phenomenon of Bose-Einstein stimulation (see \citep{Niemeyer:2019aqm}) with
\begin{equation}
     \mathcal{N} = \frac{h^3 n}{V_p} = \frac{6\pi^2 \hbar^3 n}{m^3 v^3}.
\end{equation}
Correspondingly, the scattering time is given by \citep{Levkov:2018kau}
\begin{equation}\label{eq:tau}
    \tau \simeq \frac{mv^6}{6\sqrt{2}\pi^3\hbar^3 G^2n^2\log\Lambda},
\end{equation}
where the momentum-transfer cross section $\sigma_g$ for Rutherford scattering is given by $\sigma_g\simeq \pi G^2 m^2v^{-4}\log\Lambda$, with $\Lambda\simeq R/\lambda_{\rm dB}$. 

This implies that over period of order $O(\tau)$ we may expect effect due to the quantum nature of the scalar field. One of such effect, which we shall discuss in more detail later, is the formation of solitonic structures through the Bose-Einstein condensation. 

\subsection{Soliton solutions}\label{sec:sol}

If we set the scale factor $a = 1$ and assuming $\rho\gg \langle\rho \rangle$\footnote{Normalizing $a = 1$ we are taking periods of evolution of the system that are not very large compared to the period of evolution of the universe. On the other hand, the condition $\rho\gg \langle\rho\rangle$ would be met for virialized structures that form in the post-inflationary universe.}, the SP system of equations \eqref{eq:sp} admits solutions of the form
\begin{equation}
    \psi(\textbf{x},t) = \phi(r)e^{iE t/\hbar},
\end{equation}
where $r$ is the radial coordinate and $E$ is the energy associated to the configuration. The system described by the SP equations \eqref{eq:sp} and the above ansatz has numerous solutions satisfying appropriate initial and boundary conditions \citep{Guzman:2004wj,Guzman:2006yc}. These solutions, often referred to as Newtonian boson stars (NBS), are characterized by the number of nodes present in $\psi$ before the solution asymptotically decays. The solution without nodes, the soliton solution, is the ground state of the SP system, presenting the lowest energy. Accordingly, solutions with nodes are called the excited NBSs.

The soliton solution is the  most widely studied in the literature (see for example \citep{Alvarez-Rios:2022qah,PhysRevD.69.124033,Guzmán_2006,PhysRevD.84.043531,PhysRev.187.1767,Seidel:1991zh,PhysRevLett.72.2516,Alcubierre:2001ea}). This is because the soliton is the attractor solution of the SP system: scalar field configurations with arbitrary initial conditions tend to migrate through a ``gravitational cooling" mechanism to the ground state solution of the SP system (see \citep{Seidel:1993zk}). In addition, it has been also shown in \citep{Levkov:2018kau} that initially homogeneus scalar fields with Gaussian-distributed initial conditions in momenta evolve to form localized soliton profiles by Bose-Einstein condensation in a timescale $t\sim \tau$. After its formation, a soliton accretes mass according to
\begin{equation}
    M_{\rm sol}(t)\simeq M_{\rm sol,0}\left(\frac{t}{\tau}\right)^{1/2}.
\end{equation}

 \noindent Note that the condensation time $\tau$ dictates the timescale at which the soliton structures form and evolve.
 
Once a soliton structure is virialized, its properties are mostly related to its mass $M_{\rm sol}$. For example, the half-mass radius $R_{1/2}$ and the virial velocity $v_{\rm vir}$ are given by (see for example appendix B in Ref.~\citep{PhysRevD.95.043541}): 
\begin{equation}\label{R_and_v}
    R_{1/2}\simeq \frac{4\hbar^2}{GM_{\rm sol}m^2}, \ \ \ v_{\rm vir}^2\simeq 0.4\frac{GM_{\rm sol}}{R_{\rm 1/2}}.
\end{equation}

\noindent Additionally, the coherent length $\lambda_{\rm dB}$ for the virial velocity is:
\begin{equation}
    \lambda_{\rm dB} = \frac{\hbar}{mv_{\rm vir}} \simeq 0.8 R_{1/2},
\end{equation}
which implies that the solitons formed from an scalar field present sizes of the order of the de Broglie wavelength associated to the configuration.

From the QHD system, Eq.~\eqref{eq:qhd}, the physical nature of the soliton profile can be clearly understood. Since soliton solution is a static configuration, $\partial_t\textbf{v} = 0 = \textbf{v}$. If we set for simplicity $\nabla\sim 1/\mathcal{R}_{\rm sol}$, where $\mathcal{R}_{\rm sol}$ is a characteristic radius of the soliton profile (typically~$\lambda_{\rm dB}$), we obtain,
\begin{equation}
    \nabla Q\sim -\frac{\hbar^2}{2m^2 \mathcal{R}_{\rm sol}^3}, \ \ \ \ \ \nabla \Psi\sim \frac{G M_{\rm sol}}{\mathcal{R}_{\rm sol}^2}.
\end{equation}
And the condition of equilibrium configuration yields,
\begin{subequations}
\begin{equation}\label{eq:nvn_nq}
    \nabla \Psi = -\nabla Q \ \ \Rightarrow \ \ \nu\frac{GM_{\rm sol}}{\mathcal{R}_{\rm sol}} = \frac{\hbar^2}{2m^2\mathcal{R}_{\rm sol}^2},
\end{equation}
\begin{equation*}
 \mathrm{and}\qquad   \frac{\partial\rho}{\partial t} = 0.
\end{equation*}
\end{subequations}
In the above expression $\nu$ is a dimensionless constant. 
 %\JCH{No se entiende esta frase:} In the above expression $\nu$ is a constant introduced to apply correctly the summation in the above expression (i.e. we are consdering, for example, that $\nabla Q\simeq \rm{const.} \hbar^2/(2m^2\mathcal{R}_{\rm sol}^3)$). \JCH{$\nu$ es sólo una constante numérica?}\lp{Sí, es sólo una constant. Lo que pasa es que al hacer $\nabla\Psi = -\nabla Q$ se tendría (al aproximar $\nabla = Cte/\mathcal{R}_\rm{sol}$ y no sólo $\nabla \sim 1/\mathcal{R}_\rm{cal}$) $C_1\frac{GM_{\rm sol}}{\mathcal{R}_{\rm sol}} = C_2 \frac{\hbar^2}{2m^2\mathcal{R}_{\rm sol}^{2}}$. Ya la $\nu = C_{1}/C_{2}$.}

From the above expression we can see then that the soliton profile can be understood as a result of the equilibrium between the forces generated by the quantum and the gravitational potentials. Using Eq.~\eqref{eq:nvn_nq} it is also evident that the soliton profile must fulfill the condition
\begin{equation}
    \mathcal{R}_{\rm sol} = \frac{1}{2\nu}\frac{\hbar^2}{GM_{\rm sol}m^2},
\end{equation}
which mantains the same parameter dependence found in the exact numerical treatment, Eq.~\eqref{R_and_v}.

\begin{figure}[H]
    \centering
    \includegraphics[width=3.4in]{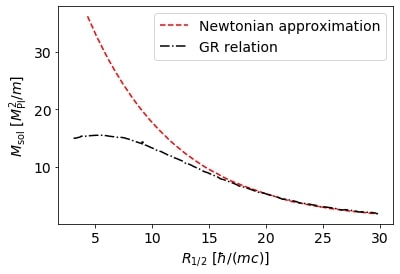}
    \caption{\footnotesize{Mass-radius relation for the soliton profile obtained from the Newtonian approximation, Eq.~\eqref{R_and_v}, and the General Relativistic treatment.}}
    \label{fig:mrr}
\end{figure}

\textit{A general-relativistic regime.-} We can anticipate from the relation $M_{\rm sol}\propto R_{1/2}^{-1}$ (from Eq.~\eqref{R_and_v}) that for certain soliton masses we should expect that a general relativistic treatment must be necessary to describe the configurations adequately. In fact, when general relativistic effects are incorporated to the system a different mass-radius relation for the soliton profile is obtained for small radius (large masses)\footnote{See Ref~\citep{Guzman:2009xre} for a recent review in how to obtain the soliton solution for the complete EKG system of equations.} (see Fig.~\ref{fig:mrr}). In particular, a limiting maximum mass is predicted to exist for the soliton profile. Such critical mass has been widely studied in the literature \citep{PhysRev.187.1767,Seidel:1991zh,PhysRevLett.72.2516,Alcubierre:2001ea} and is given by:
\begin{equation}\label{eq:m_crit}
    M_{\rm sol}^{\rm (crit)}\simeq 0.633\frac{m_{\rm Pl}^2}{m}.
\end{equation}
Above this critical mass, no stable soliton solutions are expected to exist. This is because, for soliton structures with larger masses, we anticipate that the force generated by the quantum potential is insufficient to balance the gravitational force resulting from the self-gravity of the system. This phenomenon is equivalent to the Chandrasekhar mass limit for white dwarf stars but associated to soliton structures. 

In the case of configurations that include excited states of a scalar field with a specific mass, it has been demonstrated that the resulting configurations can possess larger masses \citep{Seidel-Suen1990,Hawley2003,Urena2009,bernal2010multistates,
urena-bernal1}. However, as previously discussed, these excited states have been found to be unstable and undergo gravitational cooling, ultimately transitioning to the ground state solution.

\subsection{The Schr\"odinger-Vlasov correspondence}\label{sec:spc}

As mentioned earlier, we can use either the SP or the QHD equations to explain the process of structure formation during the slow-reheating epoch. However, this approach necessitates addressing the characteristic length scale $\lambda_{\rm dB}$, which is significantly much smaller than the cosmological horizon in the post-inflationary universe. Consequently, modeling the gravitational collapse of a structure using either of the previously discussed formulations becomes a challenging task in general.

Nonetheless, there exists an alternative approach, although it is approximate in nature, which allows us to simplify the resolution of the $\lambda_{\rm dB}$ scale and capture the significant effects arising from the wave mechanics of the field on large scales (much larger than $\lambda_{\rm dB}$). This method involves smoothing out the intricate details of the dynamics occurring at small scales (less than or approximately equal to $\lambda_{\rm dB}$) governed by the SP equations.  We shall elaborate more on the details of this description in this section.

Based on the studies conducted in \citep{Dawoodbhoy:2021beb,PhysRevA.40.2894,1993ApJ...416L..71W,PhysRevD.97.083519,PhysRevD.96.123532} we choose to utilize the Husimi representation of $\psi$, which is a smoothed phase space representation. This representation, introduced in \citep{1940264}, involves smoothing out $\psi$ using a Gaussian window with a width parameter $\eta$ and subsequently performing a Fourier transform of the form
\begin{eqnarray}
    \tilde\Psi(\textbf{x,p},t) &=& \frac{1}{(2\pi\hbar)^{3/2}}\frac{1}{(\eta\sqrt{\pi})^{3/2}}\nonumber \\
    &\times & \int e^{-\frac{(\textbf{x}-\textbf{y})^2}{2\eta^2}}\psi(\textbf{y},t)e^{-i\frac{\textbf{p}\cdot(\textbf{y}-\textbf{x}/2)}{\hbar}}d^3\textbf{y}.
\end{eqnarray}
The Husimi distribution function, which defines an smoothed mass density structure of phase space, is defined as
\begin{equation}
    \mathcal{F}(\textbf{x},\textbf{p},t)=|\tilde\Psi(\textbf{x},\textbf{p},t)|^2.
\end{equation}
The methodology for handling this distribution closely mirrors the techniques used in conventional Wigner function analysis. However, in this context, we replace the Wigner function with the newly introduced distribution function $\mathcal{F}$. Consequently, we can apply the same methods as before to derive local number density, bulk velocity, and velocity dispersion.

Furthermore, we can calculate the equation of motion for $\mathcal{F}$ in a manner akin to the Wigner-Moyal equation. This involves computing $\partial\mathcal{F}/\partial t$ and then replacing it into the SP equation. The resulting equation, when smoothed over scales significantly larger than $\lambda_{\rm dB}$ ($\eta \gg \lambda_{\rm dB}$), simplifies to the CBE or Vlasov equation:
\begin{equation}\label{eq:boltz}
    \frac{d\mathcal{F}}{dt} = \frac{\partial \mathcal{F}}{\partial t}+\frac{p_i}{m}\frac{\partial\mathcal{F}}{\partial x_i}-\frac{\partial \Psi}{\partial x_i}\frac{\partial\mathcal{F}}{\partial p_i} = 0.
\end{equation}
This last equation is the same equation that is typically used to describe the structure formation process for a dust-like component, as it is the case, for example, of the CDM model for dark matter. The only difference is that in the case of dust, we would replace $\mathcal{F}$ with the phase space distribution function of its collisionless N-body particles. Other than that, both formalisms are entirely equivalent, implying that at scales larger than $\lambda_{\rm dB}$, the dynamics of dust and the dynamics of a scalar field must be entirely identical.

\subsubsection{The fluid approximation}

Similarly than in Sec.~\ref{sec:PSF}, it is also possible to find fluid equations from this description by computing momentum moments of the CBE. Such procedure is well explained in \citep{Dawoodbhoy:2021beb}, which the reader should consult for a more comprehensive discussion. However, in particular the $0^{\rm th}$ and $1^{\rm st}$ momentum of Eq.~\eqref{eq:boltz} are found to mimic the continuity and momentum equations of classical hydrodynamics
\begin{subequations}\label{Eq:Dispersion-sys}
\begin{equation}
    \frac{\partial \rho}{\partial t}+\frac{\partial (\rho v_i)}{\partial x_j} = 0,
\end{equation}
\begin{equation}
    \frac{\partial v_i}{\partial t}+v_j\frac{\partial v_i}{\partial x_j} + \frac{1}{\rho}\frac{\partial P_{ij}}{\partial x_j}+\frac{1}{m}\frac{\partial \Psi}{\partial x_i} = 0.
\end{equation}
\end{subequations}
The quantity $P_{ij}$ is defined as $P_{ij}\equiv\rho\sigma_{ij}^2$, where $\sigma_{ij}^2$ represents the phase space velocity dispersion. Remarkably, this $P_{ij}$ serves as an effective ``pressure" term, analogous to the quantum pressure tensor $\Pi_{ij}$ found in the exact QHD equations. 

Comparing the results from Section~\ref{sec:PSF} with the system in Eq.~\eqref{Eq:Dispersion-sys}, we note that the process of smoothing over scales much greater than $\lambda_{\rm dB}$ reduces the ``quantum pressure" tensor $\Pi_{ij}$ to an effective velocity dispersion tensor  $P_{ij}$  derived from the CBE. Thus, this tensor accounts for the effects of the quantum potential/pressure on large scales. This velocity dispersion plays a crucial role in the process of structure formation, as it is essential for the stability of galactic systems, opposing the gravitational collapse and maintaining dynamical equilibrium in the system. Clearly, a small velocity dispersion can result in collapse, while excessive dispersion may cause the system to disintegrate. In summary, the effective pressure resulting from the velocity dispersion, generated by the quantum potential of the scalar field, may act to prevent PBH formation at scales above the characteristic de Broglie wavelength. Such effects were investigated in detail in Ref.~\citep{Padilla:2021zgm} (see also \citep{Harada:2022xjp}). In a subsequent section, we describe the necessary conditions under which the collapse of inflaton perturbations can indeed take place and form PBHs.

\subsection{Soliton cores and its halo-like exterior}

\begin{figure}
    \centering
    \includegraphics[width=3.5in]{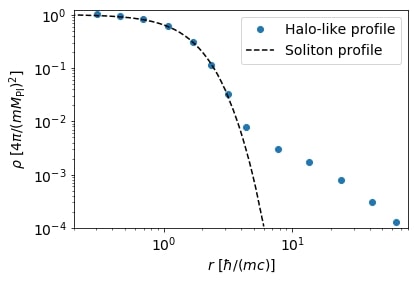}
    \caption{\footnotesize{In dotted blue is plotted a halo-like structure that is formed in cosmological simulations whereas in dashed black is plotted the soliton profile, predicted by the theory reviewed in Sec.~\ref{sec:sol}.}}
    \label{fig:core_halo}
\end{figure}

In the context of dark matter, the first realistic simulations of a scalar field considering cosmological initial conditions were conducted by Refs.~\citep{2014NatPh..10..496S,PhysRevLett.113.261302}. In that work, the authors solved the SP system of equations, Eqs.\eqref{eq:sp}, considering a scalar field as the only constituent of the universe. One of their key findings was that the final structures formed from this cosmological scalar field can be well-described by an inner soliton profile (described by the theory we reviewed in Sec.~\ref{sec:sol}) surrounded by an NFW-like envelope from a radius determined by the incoherent fluctuations of the scalar field\footnote{Note that this configuration aligns completely with the Schrödinger-Vlasov correspondence discussed in Sec.~\ref{sec:spc}; if we smooth the profile derived from simulations over scales larger than $\lambda_{\rm dB}$, we should obtain a final distribution similar to that of an NFW profile, typically obtained in the conventional dust-like scenario.}. A schematic plot of such configurations is shown in Fig.~\ref{fig:core_halo}. This soliton-envelope  structure has been confirmed to exist by several studies that considered simpler scenarios \citep{Schwabe:2016rze,PhysRevD.94.123523,Alvarez-Rios:2023cch}, e.g., in the nearly simultaneous merger of several soliton configurations \citep{mocz2017galaxy}. In the context of reheating, such core-envelope structures have also been reported \citep{Eggemeier:2021smj}.

Numerical simulations prescribe a soliton-halo\footnote{For the sake of clarity, we use here the term `halo' to define the composite structure consisting of the central soliton together with the NFW-like envelope.} mass relation given by $M_{\rm sol}\propto M_{\rm halo}^{1/3}/m$. Several hypotheses have been proposed to justify this relation. For instance, it has been suggested that the specific energy of the central soliton and the host halo are of the same magnitude \citep{bar2018galactic}, while others suggest that their specific kinetic energy is what is comparable \citep{PhysRevD.99.103020}. There is also the suggestion that the velocities of the soliton and the host halo, such as the circular, virial, or dispersion velocity, are equivalent \citep{PhysRevD.100.083022,mocz2017galaxy}. Regardless of the interpretation, the above studies agree in the following relation:
\begin{equation}\label{mcmhrelation}
\frac{GM_{\rm sol}m}{\hbar}\simeq \sqrt{\frac{3G M_{\rm halo}}{10R_{\rm halo}}}.
\end{equation}
    
\noindent In the above expression $M_{\rm halo} = (4\pi/3)\rho_{200}(a_{\rm NL})R_{\rm halo}^3$ is the total mass of the halo structure, which should coincide with the mass of the cosmological horizon evaluated at the horizon crossing time (see Eq.~\eqref{mass}), 
%\begin{equation}\label{eq:mass_h}
    %M_{\rm halo} = \frac{m_{\rm Pl}^2}{2H_{\rm end}}\left(\frac{a_{\rm HC}}{a_{\rm end}}\right)^{3/2}(1+\delta_{\rm end}(k)),
%\end{equation}
$R_{\rm halo}$ is the virial halo radius, $\rho_{200}(a_{\rm NL}) = 200\rho_b(a_{\rm NL})$, and subindex $_{\rm NL}$ refers to quantities evaluated at the time the $k$-mode becomes non-linear\footnote{\label{footnote6}The number of $e$-folds after horizon crossing necessary for a perturbation to become nonlinear can be simply expressed as $\Delta N_{\rm NL}(k) \equiv N_{\rm NL}(k)-N_{\rm HC}(k) = \ln(1.39\delta_{\rm HC}^{-1}(k))$ \citep{Padilla:2021zgm}.  The discrepancy between this value and Eq.~\eqref{eq:N_c} is less than one e-fold of expansion. In this paper we will interchangeably use both quantities.}. We can re-express Eq.~\eqref{mcmhrelation} more conveniently as follows:
\begin{equation}
    \left(\frac{M_{\rm sol}(k)}{2.4\times 10^{-5}~\rm{g}}\right) \simeq \frac{\rho_{11}^{1/6}(a_{\rm NL})}{m_5}\left(\frac{M_{\rm halo}(k)}{7.1\times 10^{-2}~\rm{g}}\right)^{1/3}, \label{mc_mh}
\end{equation}
Or equivalently, with some more algebra,
\begin{equation}\label{m_sol_halo}
M_{\rm sol}(k)\simeq 10\,\frac{\sqrt{\delta_{\rm HC}(k)}}{m_5}~\rm{g}\,.
\end{equation}
In the above expressions $m_5\equiv m/(10^{-5}~m_{\rm Pl})$ and $\rho_{11}(a)\equiv \rho_{200}(a)/(10^{11}~\rm{GeV})^4$. This is an important result since, for a given $m_5$, we anticipate that the mass of the solitons formed in the post-inflationary universe (during reheating) must be closely related to the amplitude of the perturbations $\delta_{\rm HC}(k)$ and independent of the mass of its host halo. For example, in the case where $m_5 = 1$ and $\delta_{\rm HC}\sim 10^{-5}$ (typically predicted by CMB observations) we have that the solitons formed in this context should have a mass of $M_{\rm sol}(k)\simeq 6.54\times 10^{-4}~\rm{g}$.

%Assuming the validity of Eq. (50) using $\tau$ from Eq. (35) as a parameterization of soliton mass growth (despite lacking theoretical support), the explanation for Eq. (51) lies in the saturation of mass growth [63]. Initially governed by the virial temperature of the halo, the ambient boson field surrounding the star reaches a point where the soliton's own virial temperature becomes dominant, causing a slowdown in mass growth. This saturation occurs when the two temperatures align, resulting in a stabilized soliton mass. Ongoing condensation contributes to continued, albeit reduced, mass growth.

\section{PBH formation during slow-reheating:  Collapse from primordial structures}\label{Sec:Slow-reheat-phbs}

\subsection{New criteria of PBH formation in a slow-reheating scenario}\label{Sec:VI.2}

As we previously shown, if reheating lasts long enough, two types of structures could form during this phase. Firstly, the formation of halo-like structures that resemble a typical NFW profile when smoothing over scales larger than $\lambda_{\rm dB}$. On the other hand, when considering scales $R\sim \lambda_{\rm dB}$, the formation of soliton-like structures is expected. In this section, we will investigate the conditions under which we can expect the gravitational collapse of both types of structures onto PBHs, closely following Ref.~\citep{Padilla:2021zgm}.

\subsubsection{Halo collapse}\label{sec:halo_collapse}

If the halo-like structures that arise following inflation are massive enough, they could become gravitationally unstable and collapse, resulting in the formation of PBHs. Specifically, a reliable indicator of such collapse is when overdensities reach a state of virialization within a radius for the associated halo comparable to, or smaller than, the Schwarzschild radius; that is, $R_{\rm Sch} \equiv 2GM_{\rm halo}$. In other words, if $R_{\rm halo}\leq R_{\rm Sch}$, the collapse to PBHs is likely to occur. By comparing the halo's virial  radius with the Schwarzschild radius, we can determine that PBH formation will always take place when the following condition is satisfied:
\begin{equation}\label{eq:M_c_h}
M_{\rm halo}\geq \frac{3.144\times 10^{34}}{\sqrt{\rho_{11}(a_{\rm NL})}}~\rm{GeV}.
\end{equation}
When this inequality is combined with Eq.~\eqref{eq:be} and the mass of the halo (which coincides with the mass of the cosmological horizon at the horizon crossing time), it reduces to the following condition:
\begin{equation}
\Delta N_{\rm NL}(k)\equiv N_{\rm NL}(k)-N_{\rm HC}(k) \leq \frac{2}{3}\ln[14.14].
\end{equation}
By using the relationship $\Delta N_{\rm NL}(k) = \ln[1.39\delta_{\rm HC}^{-1}(k))]$  (see footnote \ref{footnote6}), we can finally obtain the condition for PBH formation:
\begin{equation}\label{eq:th_halo}
\delta_{\rm HC}(k)\geq \delta_{\rm th}^{\rm (halo)}\equiv 0.238.
\end{equation}

\subsubsection{Soliton collapse}

In accordance with our previous discussion in Sec. \ref{sec:sol}, there is a maximum possible mass for soliton structures, as described by Eq.~\eqref{eq:m_crit}. This maximum mass is reached when the quantum pressure resulting from the Heisenberg uncertainty principle is insufficient to counterbalance the self-gravitational forces within the soliton structure. By substituting this maximum mass into Eq.~\eqref{m_sol_halo}, we can determine a critical threashold value $\delta_{\rm th}^{\rm (soliton)}$ beyond which the central soliton becomes unstable and may collapse to form a PBH. The condition for PBH formation in this scenario is given by\footnote{In the context of the SFDM, this mechanism has been also proposed to explain the formation of supermassive black holes in the model (see for example \citep{Padilla:2020sjy}).}:
\begin{equation}\label{eq:th_soliton}
\delta_{\rm HC}(k)\geq \delta_{\rm th}^{\rm (soliton)}\equiv 0.019.
\end{equation}
\subsection{Overview of the mechanisms of PBH formation in slow-reheating}

\begin{figure}[H]
    \centering
    \includegraphics[width=3.5in]{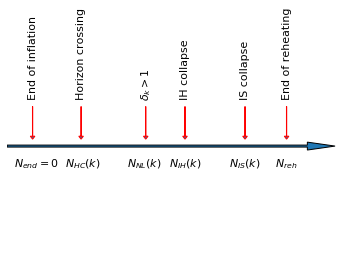}
    \caption{\footnotesize{Evolutionary stages of density fluctuations during the reheating period as a function of the number of $e$-folds after inflation concludes ($N_{\rm end}$). After the onset of reheating, a mode with a specific wavenumber $k$ reenters the horizon $N_{\rm HC}(k)$ $e$-folds later, reaching a nonlinear amplitude at $N_{\rm NL}(k)$. This nonlinear state gives rise to the formation of a halo-like structure (or a PBH) after a Hubble time, at $N_{\rm halo}(k)$. Once the halo undergoes virialization, the condensation of the inflaton at its core generates a soliton-like structure (or a PBH), emerging at $N_{\rm soliton}(k)$. Reheating is expected to reach completion and achieve thermalization at $N_{\rm reh}$. However, if this final event occurs earlier, the sequential progression is disrupted, and only some or none of the $k$-dependent processes may occur.}}
    \label{fig:PBH_timeline}
\end{figure}

Let us describe the integral picture of the variety of mechanisms leading to the formation of PBHs during reheating. Fig.~\ref{fig:PBH_timeline} summarizes this general timeline, which we will discuss in more detail below.

As previously stated, during the slow-reheating phase we expect a few perturbation modes (the ones at the smallest scales in the spectrum) to reenter the horizon. The number of $e$-folds at horizon reentry $N_{\rm HC}(k)$, after inflation ends, can be calculated using the following equation:
\begin{equation}\label{eq:hc}
N_{\rm HC}(k) = 2\ln\left( \frac{k_{\rm end}}{k}\right).
\end{equation}

Once perturbations reenter the horizon, they grow as $\delta\sim a$ (see Eq.~\eqref{delta_a}) and may reach a nonlinear stage. The number of $e$-folds necessary to reach this regime depends on the wave number $k$ and the density contrast amplitude at horizon crossing $\delta_{\rm HC}(k)$, and it can be expressed as  (see footnote \ref{footnote6}):
\begin{equation}
N_{\rm NL}(k) = N_{\rm HC}(k)+\ln[1.39\delta_{\rm HC}^{-1}(k)].
\end{equation}

When inhomogeneities reach a nonlinear amplitude, halo-like structures or PBHs are expected to form within a Hubble time. Expressed in terms of $e$-folds, this happens at:
\begin{equation}\label{eq:N_halo}
N_{\rm halo}(k) = N_{\rm NL}(k)+\frac{2}{3}\ln\left(1+\frac{H^{-1}}{t_{\rm NL}(k)}\right),
\end{equation}
where $t_{\rm NL}(k) = [2/(3H_{\rm end})][e^{N_{\rm HC}(k)}1.39/\delta_{\rm HC}(k)]^{3/2}$. The collapse time (or the number of $e$-folds up to collapse) approximately corresponds to the time derived in Eq.~\eqref{eq:N_c}, indicating that this collapse criterion is necessary but not sufficient for PBH formation.

\begin{figure}[H]
    \centering
    \includegraphics[width=3.5in]{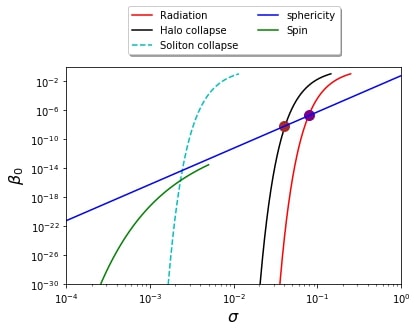}
    \caption{\footnotesize{The PBH density fraction $\beta_0$ plotted as a function of the variance of the density contrast $\sigma$. The blue and green lines are plotted by following Eqs.~\eqref{sigma1} and \eqref{sigma3}, respectively. The solid black and dashed cian lines were plotted by using the standard Press-Schechter formalism, Eq.~\eqref{eq:ps}, and the threashold values \eqref{eq:th_halo} and \eqref{eq:th_soliton}, respectively. For comparison, we also included the PBH density fraction $\beta_0$ calculated in a standard radiation-dominated universe, where we took $\delta_{\rm th}^{\rm (rad)} =0.41$ (see for example \citep{Harada:2013epa}). In the plot we have also included the brown and purple circles, which denotes the points at which the sphericity criterion becomes more restrictive for PBH formation (for larger $\sigma$) than the criterion related to velocity dispersion or the case in which the PBH formation happens in a radiation-dominated universe, respectively.}}
    \label{fig:beta_sigma}
\end{figure}

During the collapse, three important effects may hinder the gravitational collapse into PBHs. To wit, the sphericity of the configurations, the conservation of angular momentum, and the velocity dispersion. The first two criteria limit the abundance of PBHs as a function of the variance of fluctuations (as discussed in Section~\ref{sec4}). The latter effect can be expressed in terms of a threshold amplitude for collapse, as shown in the previous section. With the aid of the Press-Schechter integral, we can express the abundance at the time of formation, $\beta_0$, in terms of the variance by assuming a Gaussian probability density. We thus bring all these effects together in Fig.~\ref{fig:beta_sigma}, which shows limits to the abundance at the time of formation, $\beta_0$, that should be carefully interpreted. In the consideration of direct collapse, the velocity dispersion criterion (the halo collapse/black curve), seems to impose the most stringent bound to the production of PBHs. However, if configurations do not virialize, one could follow the evolution of fluctuations as that of dust, which subjects the abundance of collapsed objects to the sphericity (blue line) criterion at larger variance values. Note that the distribution of spin in the initial fluctuations yields the limit imposed by the green line, which is never the most stringent bound to the production of PBHs.

The above is, however, not the full story. If reheating extends sufficiently to reach $t_{\rm soliton}(k) = t_{\rm NL}(k)+\tau(k)$, a soliton-like structure gets to form at the core of virialized haloes. Then, a new type of PBHs may emerge at the center of the halo-like structures via the collapse of the Bose-Einstein condensate. The condensation process initiates when the inhomogeneity becomes nonlinear, and it can be described by the following equation:
\begin{equation}
\frac{\tau(k)}{t_{\rm NL}(k)} =8.168\times 10^{-18}(m_5^2M_{\rm Pl}^2M_{\rm halo}(k)R_{\rm halo}(k))^{3/2}
\end{equation}
In the above expression, we use the condensation time $\tau$ from Eq.~\eqref{eq:tau} and reexpress it in terms of halo quantities. The condensation time thus determines the number of $e$-folds necessary for soliton/PBH structures to form:
\begin{equation}\label{eq:N_soliton}
N_{\rm soliton}(k) = N_{\rm NL}(k)+\frac{2}{3}\ln\left(1+\frac{\tau(k)}{t_{\rm NL}(k)}\right).
\end{equation}
The criterion for discriminating PBHs from solitons is given by the threshold value of the overdensity at horizon crossing, presented in Eq.~\eqref{eq:th_soliton}. The associated abundance as a function of the variance is presented in Fig.~\ref{fig:beta_sigma} by the dashed line. This is an alternative route to the direct collapse and the abundance is thus not subject to the sphericity or spin criteria. Let us emphasize that if reheating is terminated early enough, the soliton collapse will not take place.

In concluding this section, we highlight a final possibility, not been previously mentioned in this paper. In the process of formation of structures, such as halos or solitons, with a mass below their critical collapse threshold, the process of accreting matter from their surroundings can be acheived if reheating lasts long enough. Consequently, the structures can grow in mass until they reach the critical value required for them to collapse and the formation of black holes. This particular possibility has been explored in Ref.~\citep{DeLuca:2021pls}. Illustrating this mechanism in a plot equivalent to Fig.~\ref{fig:beta_sigma} is a task to be tackled in future work. 
%\JCH{Es esto último cierto? Aún no se puede graficar en la fig. 6 este caso????}\lp{Pues de hecho todavía no se puede pero creo que no sería taaan dificil obtenerlo. La idea sería que tendríamos una $\delta_{th}$ más chica a la que derivamos y que estaría como función del tiempo. Podríamos tener $\delta_{th}$ más chicas si son escalas $k$'s que formaron inflaton-halos a tiempos más tempranos (ya que tendrían más tiempo de acretar materia hasta alcanzar la masa crítica de colapso). Creo que podría ser un ejercicio que podría sacar Tadeo y quizá de ahí podría salir algo interesante. Digo, creo que al final sería un criterio más completo para obtener el threshold para el colapso que el que propusimos nosotros.}

\section{Testing the mechanisms in a simple setting}
\label{sec:example}

\begin{figure}[H]
    \centering
    \includegraphics[width=3.5in]{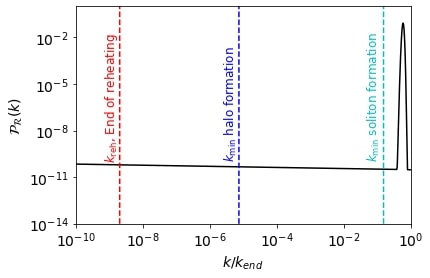}
    \caption{\footnotesize{Power spectrum as a function of the scale wavenumber $k$. The red, blue, and cyan lines represent the minimum $k$-modes relevant to the reheating epoch, the formation of halo-like structures, and the formation of solitonic structures. It's worth noting that this figure bears a striking resemblance to Fig.~2 in \citep{Hidalgo:2022yed}, with the sole distinction being that it considers a reheating period lasting for $N_{\rm reh} = 30$ $e$-folds of expansion, as opposed to the 40 $e$-folds used in \citep{Hidalgo:2022yed}.}}
    \label{fig:pps}
\end{figure}

Let us apply the presented hypothesis to a relevant inflationary scenario. We consider a power spectrum parameterized by the following expression
\begin{equation}\label{eq:power_spectrum}
    \mathcal{P}_{\mathcal{R}}(k)=\mathcal{A}_s\left(\frac{k}{k_{*}}\right)^{n_{s}-1}+\mathcal{B}_{s}\exp\left[-\frac{(k-k_{p})}{2\Sigma_{p}^{2}}\right],
\end{equation}
%\JCH{This is not an equation}
where $k_* = 0.05~\rm{Mpc^{-1}}$ is a pivot scale. We thus approximate the power spectrum with the usual slow-roll approximation plus a Gaussian peak located at $k_p$ and with a variance $\Sigma_p^2$. For this example we shall consider the model parameters $\mathcal{A}_s = 2.099\times 10^{-9}$, $n_s = 0.9634$, $\mathcal{B}_s = 0.084$, $\Sigma_p = 0.03 k_{\rm end}$, $k_p = 0.6 k_{\rm end}$, and $k_{\rm end} = 0.346~\rm{m^{-1}}$. The power spectrum generated for this set of parameters is sketched in Fig.~\ref{fig:pps}.

From the bound in the tensor-to-scalar ration ($r\leq 0.032$) \citep{tristram2022improved}, we can impose a constraint on the Hubble parameter evaluated at the horizon crossing time of the pivot scale $k_*$:
\begin{equation}
    H_* = \sqrt{\frac{\mathcal{A}_s r}{2}\pi M_{\rm Pl}}\leq 4.44\times 10^{13}~\rm{GeV}. 
\end{equation}
Assuming that $H_*\gtrsim H_{\rm end}$ and using the Friedmann equation, we limit the energy scale at witch the end of inflation took place:
\begin{equation}
    \rho_{\rm end} \leq (1.368\times 10^{16}~\rm{GeV})^4.
\end{equation}
For definiteness we adopt the maximum value allowed, $\rho_{\rm end} = (1.368\times 10^{16}~\rm{GeV})^4$. 

We turn our attention to the value of the  density contrast $\delta_{\rm HC}(k)$ evaluated at the horizon crossing time. If we consider the mean amplitude of density perturbations as that of the configurations to collapse, taken as, 
\begin{equation}
    \bar\delta_{\rm HC}(k) = \left(\frac{\delta \rho}{\rho}\right)_{k=aH} = \sqrt{\mathcal{P}_\delta(k)},
\end{equation}
\noindent
we can compute the number of $e$-folds necessary for each particular scale to reenter the cosmological horizon, Eq.~\eqref{eq:hc}, then form halo-like structures, Eq.~\eqref{eq:N_halo}, and finally form soliton-like structures, Eq.~\eqref{eq:N_soliton}.\footnote{In a more accurate approximation, $\delta_{\rm HC}(k)$ follows a Gaussian distribution, with a specific number of $e$-folds required for the collapse of each amplitude. By taking the period required for collapse from the mean amplitude, we are  underestimating the the mass of primordial structures and overestimating the collapse time of the tail of the distribution.} %\JCH{Underestimating???}\lp{De hecho acá no estoy seguro si sería sobreestimar o subestimar. Estamos asumiendo que en lugar de que $\delta$ tenga una distribución Gaussiana tendríamos que todas las perturbaciones tienen el valor de la media. ¿Eso no sobreestimaría la abundancia?}.  This approximation is thus conservative and we....\JCH{Acá depende de lo que me respondas para ver que ponemos}.}.

%------------------------
%\JCH{Esto no va aquí y no tiene sentido ponerlo... ya la discusión de la varianza y su relación con }\lp{Al menos la relación de $\mathcal{P}_\delta$ con $\mathcal{P}_\mathcal{R}$  no la habíamos puesto. ¿No debería ir?}
%which is given by the following expression
%\begin{equation}
%    \bar\delta_{\rm HC}(k) = \left(\frac{\delta \rho}{\rho}\right)_{k=aH} = \sqrt{\mathcal{P}_\delta(k)},
%\end{equation}
%where in the above expression $\mathcal{P}_\delta(k)$ is the power spectrum of density perturbations, which is related to the power spectrum via the following relation
%\begin{equation}
%    \mathcal{P}_\delta(k,t) = \frac{4}{25}\left(\frac{k}{aH}\right)^4 \mathcal{P}_\mathcal{R}(k,t).
%\end{equation}
%\JCH{-------------------------}
 
 Having computed that, in Fig.~\ref{fig:pps} we mark the minimum $k$'s (largest scales) that undergo each of these three processes. Let us stress that we have assumed  a reheating period that lasted for 30 $e$-folds of expansion. 

\begin{figure}[H]
    \centering
    \includegraphics[width=3.5in]{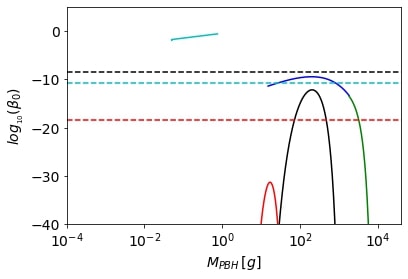}
    \caption{\footnotesize{The density fraction $\beta_0$ as a function of $M_{\rm PBH}$ for the collapse of solitons (cian) and halo-like structures (black). We have also included the sphericity criterion (blue) and the conservation of angular momentum criterion (green). For comparison, we have included the abundance of PBHs formed in a radiation-dominated universe. We also showed the current constraints on the abundance of PBHs for each of the scenarios studied and the mass range of PBHs formed. In particular, the dashed cian, black, and red lines are constraints that apply for PBHs that form due to the collapse of solitons, halos, and perturbations during a radiation-dominated epoch, respectively (see main text for further discussion).}}
    \label{fig:enter-label}
\end{figure}

For each one of the scales that manage to form halos or solitons, we compute the probability of PBH formation. For this purpose, we employ  the Press-Schechter formalism (see appendix \eqref{AP:PS}), where we calculate the variance of the contrast density following Eq.~\eqref{eq:variance} and use the threshold values given by Eqs.~\eqref{eq:th_halo} and \eqref{eq:th_soliton}. The result obtained is shown in Fig.~\ref{fig:enter-label}. For comparison, we also included the abundance of PBH predicted by the sphericity criterion, Eq.~\eqref{sigma1}, and the conservation of angular momentum criterion, Eq.~\eqref{sigma3}, that would modulate the PBH production in the case of pure dust. As expected, the velocity dispersion criterion reduces the abundance of PBHs with respect to the other criteria.

On the other hand, we can observe that due to the small value of the threshold for collapse of soliton structures, in this example we would have a much larger abundance of PBHs due to the collapse of the central soliton of the halos. It is worth mentioning that in the case of PBHs formed by solitons we would not have a one to one relationship of scale vs. mass of the PBH. This is due to how the soliton's mass depends on the total mass of its host halo (see Eq.~\eqref{m_sol_halo}). For a more detailed discussion of this, we refer the reader to Ref.~\citep{Hidalgo:2022yed}.

It is evident that the given power spectrum may result in the formation of PBHs with different mass ranges, depending on the duration of reheating and the dominant matter content. Specifically, our analysis shows that PBHs with masses around $0.74$, $202$, or $17$ grams are more prominently produced when they originate from the collapse of scalar field solitons, scalar field halos, or overdensities in a radiation-dominated universe, respectively. Each of these PBH populations influence specific phenomena at astronomical and cosmological levels, thus meeting different constraints depending on their mass.

In Fig.~\ref{fig:enter-label}, we have included the respective constraints applicable to each scenario. To derive these constraints, we used the publicly available software PBHBeta \citep{Gomez-Aguilar:2023bej}, which enabled us to determine constraints for PBHs at various mass ranges in extended reheating scenarios. These constraints were adjusted for two cases: one where the reheating period lasts for $N_{\rm reh} = 30$ $e$-folds of expansion (to calculate the constraints in the case of collapsing halo- and soliton-like structures) and another with $N_{\rm reh} = 0$ to obtain constraints for the collapse during the radiation-dominated scenario. Interestingly enough, our analysis shows that in the case of PBH formation during a radiation epoch, the abundance of PBHs formed would be well below the limits imposed by observations. On the other hand, for PBHs formed from the collapse of Inflaton halos, the abundance is much closer to the limit value. Finally, for the same primordial power spectrum, we would observe an overabundance of PBHs resulting from the collapse of solitonic structures, surpassing the limits imposed by observations.

{To conclude this section, it is necessary to mention that the abundances obtained in this example would indicate that models with features similar to that in Eq.~\eqref{eq:power_spectrum} would be strongly ruled out in a scenario of extended reheating where reheating lasts long enough. However, in a short reheating period, where PBHs from the peak in the primordial power spectrum do not have time to form, or where the peak in the primordial power spectrum becomes smaller, the simple model \eqref{eq:power_spectrum} still satisfies the constraints imposed by the Planck mass relics and cannot be ruled out. Of course, to draw more realistic conclusions for different inflationary models existing in the literature, it would be necessary to compare such models with the criteria for PBH formation discussed here (as was done, for example, in Ref.~\citep{Padilla:2023lbv}).}

\section{Summary and Outlook}

In this article, we have reviewed the criteria for the formation of Primordial Black Holes (PBHs) in the context of a slow-reheating scenario. 
{We focus on an extended reheating stage dominated by an oscillating field in a quadratic potential.  }
Specifically, we have examined how the gravitational collapse of primordial inhomogeneities re-entering the cosmological horizon can be influenced by three significant effects: the effects related to the morphology of the initial perturbation (how spherical or non-spherical it is), the possible angular momentum it might have presented, and the effects due to velocity dispersion. We have described in detail how the latter emerges from the quantum nature of the dominating scalar field once it is averaged over scales much larger than the associated de Broglie scale. Each of these effects pescribes a bound to the abundance of PBHs, as illustrated in Fig.~\ref{fig:beta_sigma}. In particular, we have found in the case in which $\sigma \leq 0.04$ that the criterion for PBH formation due to velocity dispersion effects is more restrictive than the criteria considering the morphology of the perturbation and its angular momentum. As a result, this criterion is the most important to consider for those values of $\sigma$. On the other hand, in the case in which $\sigma > 0.04$, the most important effect preventing the collapse of perturbations into PBHs is the one related to the morphology of the system.

At scales comparable to the de Broglie wavelength of the scalar field, we have seen that the formation of a solitonic-like structure (due to the Bose-Einstein condensation process) is expected at the center of virialized configurations that form in the post-inflationary universe. In this article, we have also reviewed the necessary condition for the formation of PBHs due to the gravitational collapse of these solitons and we have calculated the abundance of PBHs that should be expected by this mechanism (which can be also seen in Fig.~\ref{fig:beta_sigma}). In the figure it is easy to see that the collapse of solitons into PBHs is much more likely to occur than in the case of the collapse of the total perturbation.

We emphasize that in order to achieve the formation of PBHs due to either of these two formation mechanisms (collapse of the total perturbation or solitonic center), reheating should last sufficiently long to allow for each of these processes to take place. The timeline for this process is presented in Fig.~\ref{fig:PBH_timeline}. Another important remark is that we have assumed the dominating scalar field during reheating to be the inflaton field itself, oscillating around the bottom of a quadratic potential. The extension to scalar fields of other nature, such as axion monodromy \citep{Silverstein:2008sg,mcallister2010gravity}, curvaton \citep{enqvist2002adiabatic,lyth2002generating,lyth2002generating} or multiple fields \citep{iacconi2023multi,Iacconi:2021ltm}, is still work in progress.

This work has been carried out as a review article aimed at guiding the reader through the different criteria for PBH formation in the context of a slow-reheating scenario. {Our study considers exclusively the case of a scalar field (inflaton or another) oscilating around the minimum of a quadratic potential and dominating the universe prior to the standard radiation domination. Extensions to (self-)interacting fields have been mentioned in Sec.~\ref{sec:preheating}, where the computation of PBH formation requires numerical analysis. This alternative reheating scenario shall thus be addressed elsewhere.}

Our intention is to provide the proof of principle and the tools for the reader to adapt the mechanism to specific models of the early universe, in order to test different inflationary models with this PBH formation and explore the detection window. To this end, we have also included a simple example that considers a primordial power spectrum with a Gaussian peak at small scales, showing the possibility of overproduction of PBHs. This is an exciting possibility that we shall study in more detail elsewhere.

\section{Acknowledgements}
 LEP and JCH acknowledge sponsorship from CONAHCyT Network Project 304001 “Estudio de campos escalares con aplicaciones
en cosmología y astrofísica”, and through grant CB-2016-282569. The work of LEP is also supported by the DGAPA-UNAM postdoctoral grants program, by CONAHCyT México under grants
A1-S-8742, 376127 and FORDECYT-PRONACES grant 490769. JCH acknowledges financial support from PAPIIT-UNAM programme Grant IG102123 “Laboratorio de Modelos y Datos (LAMOD) para
proyectos de Investigación Científica: Censos Astrofísicos". KAM is supported in
part by STFC grants ST/T000341/1 and ST/X000931/1.\\

\appendix
\section{Press-Schechter formalism}\label{AP:PS}

In the Press-Schechter formalism \citep{Press:1973iz}, the likelihood of having collapsed objects with masses greater than $M$ is analogous to the probability that a density field, after being smoothed, surpasses the threshold value $\delta_{\rm th}$:
\begin{equation}\label{Pdelta}
    \mathbb{P}[\delta>\delta_{\rm th}] = \int_{\delta_{\rm th}}^{\infty}P(\tilde\delta)d\tilde\delta.
\end{equation}

\noindent Adopting that the probability distribution function of $\delta$, $P(\delta)$, follows a Gaussian distribution, i.e.
\begin{equation}
    P(\delta) = \frac{1}{\sqrt{2\pi}\sigma(R)}\exp\left(-\frac{\delta^2}{2\sigma(R)^2}\right),
\end{equation}
with $\sigma(R)$ the variance of $\delta$ evaluated at the horizon crossing time,
\begin{equation}\label{eq:variance}
    \sigma^2(R
    ) = \int_0^\infty \tilde W^2(\tilde kR)\mathcal{P}_\delta(\tilde k,t_{\rm HC})d\ln \tilde k,
\end{equation}
$\tilde W(kR) = \exp(-k^2R^2/2)$ the Fourier transform of the window function used to smooth the density contrast over a scale $R = 1/k$, and $\mathcal{P}_\delta$ the power spectrum of density perturbations, we can rewrite Eq.~\eqref{Pdelta} as
\begin{equation}\label{Pdelta2}
    \mathbb{P}[\delta>\delta_{\rm th}] = \frac{1}{2}\text{erfc}\left(\frac{\delta_{\rm th}}{\sqrt{2}\sigma(R)}\right).
\end{equation}
In the above expression $\erfc(x) = 1-\erf(x)$ is the complementary error function. We can finally compute the abundance of PBHs of a given mass $M$ at the time of formation, $\beta_0(M)$, by using the expression
\begin{equation}\label{betaM}
    \beta_0(M) = -2M\frac{\partial R}{\partial M}\frac{\partial \mathbb{P}[\delta>\delta_{\rm th}]}{\partial R},
\end{equation}
where the factor 2 is included to fit the cloud in cloud correction.

\bibliographystyle{unsrt}
\bibliography{biblio}

\end{document}